\documentclass[final,nopreprintline,5p,times]{elsarticle}

\usepackage{amssymb}

\usepackage[nodots]{numcompress}

\usepackage{lineno}

\usepackage{bm}
\usepackage{multirow}
\usepackage{empheq}
\usepackage{amssymb}
\usepackage{dsfont}

\begin{document}
%
%
%
%

\newcommand{\mns}[1]{\textcolor{red}{#1}}
\newcommand{\mnt}[1]{\textcolor{green}{#1}}
\newcommand{\mnr}[1]{\textcolor{cyan}{#1}}
\newcommand{\linereminder}{\noindent\textcolor{red}{\rule{\linewidth}{0.3cm}}}
\newcommand{\norm}[1]{\|#1\|}
\newcommand{\R}{\mathbb{R}}
\newcommand{\matrixinside}[1]{\bm{#1}}
\renewcommand{\vec}[1]{\bm{#1}}

\begin{frontmatter}



\title{Geo-Sketcher: Rapid 3D Geological Modeling using Geological and Topographic Map Sketches}


\author{Ronan Mendonca Amorim{\upshape,} Emilio Vital Brazil{\upshape,} Faramarz Samavati {\upshape and} Mario Costa Sousa
}

\begin{abstract}
The construction of 3D geological models is an essential task in oil/gas exploration, development and production.
However, it is a cumbersome, time consuming and error prone task mainly because of the model's geometric and topological complexity.
The model's construction is usually separated into interpretation and 3D modeling, performed by different highly specialized individuals, which leads to inconsistencies and intensifies the challenges.
In addition, the creation of models following geological rules is paramount for properly depicting static and dynamic properties of oil/gas reservoirs.
In this work, we propose a sketch-based approach to expedite the creation of valid 3D geological models by mimicking how domain experts interpret geological structures, allowing creating models directly from interpretation sketches.
Our sketch-based modeler (Geo-Sketcher) is based on sketches of standard 2D topographic and geological maps, comprised of lines, symbols and annotations.
We developed a graph-based representation to enable (1) the automatic computation of the relative ages of rock series and layers; and (2) the embedding of specific geological rules directly in the sketching.
We introduce the use of Hermite-Birkhoff Radial Basis Functions to interpolate the geological map constraints, and demonstrate the capabilities of our approach with a variety of results with different levels of complexity.
\end{abstract}


\end{frontmatter}

\nolinenumbers


\section{Introduction}
\label{sec:introduction}

\begin{figure}[ht!]
    \centering
    \includegraphics[width=0.9\linewidth]{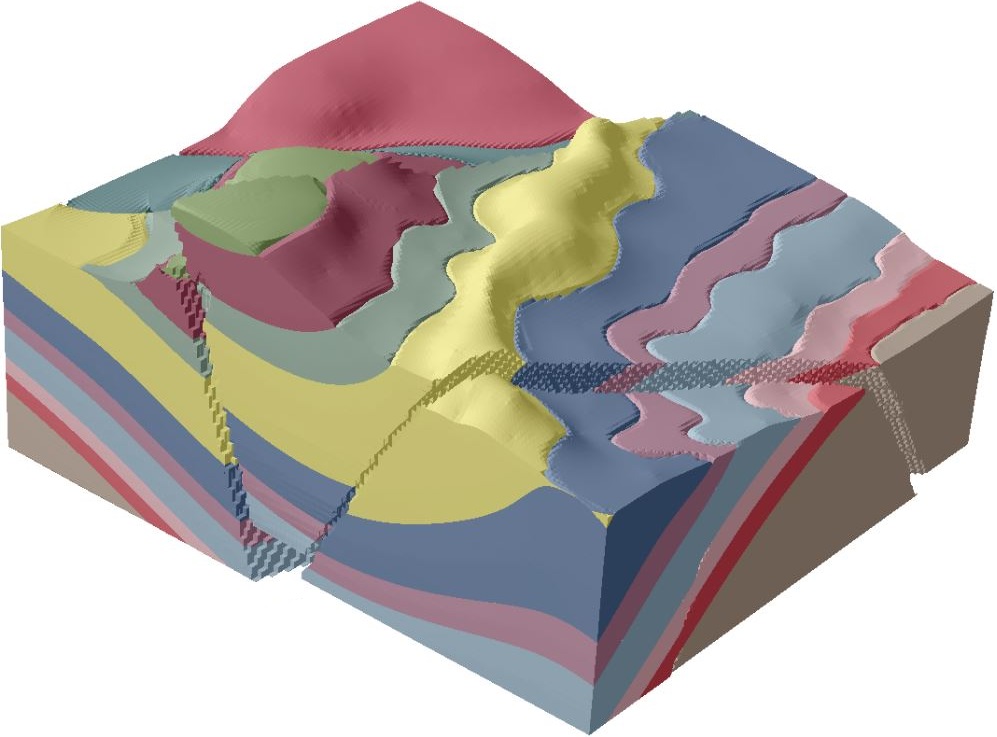}
    \caption{Example of geological model constructed using Geo-Sketcher showing complex geometry and topology.}
    \label{fig:motivation}  
\end{figure}

In this article we consider the problem of rapidly creating prototypes of 3D geological models from the geologists' 2D sketches of topographic and geological maps.
Geological models are 3D representations of the disposition, geometry and types of rocks in the subsurface.
They depict rock structures consisting of different rock layers mainly created by the deposition of sediments through periods of time.
These rock layers have been deformed by erosion and compressional/tensional forces, creating 3D solids with intricate topology and discontinuities (Fig.~\ref{fig:motivation}).

Geological models play a key role in understanding and predicting natural hazards such as earthquakes, as well as in evaluating and exploring natural resources such as ground water, minerals, and hydrocarbons (oil and gas).
In the oil and gas industry, in particular, these models are important tools for decision makers to estimate the amount of oil in-place, to decide where to drill wells, and to simulate the fluid flow behavior in a reservoir.

\begin{figure}[b!]
    \centering
    \includegraphics[width=\linewidth]{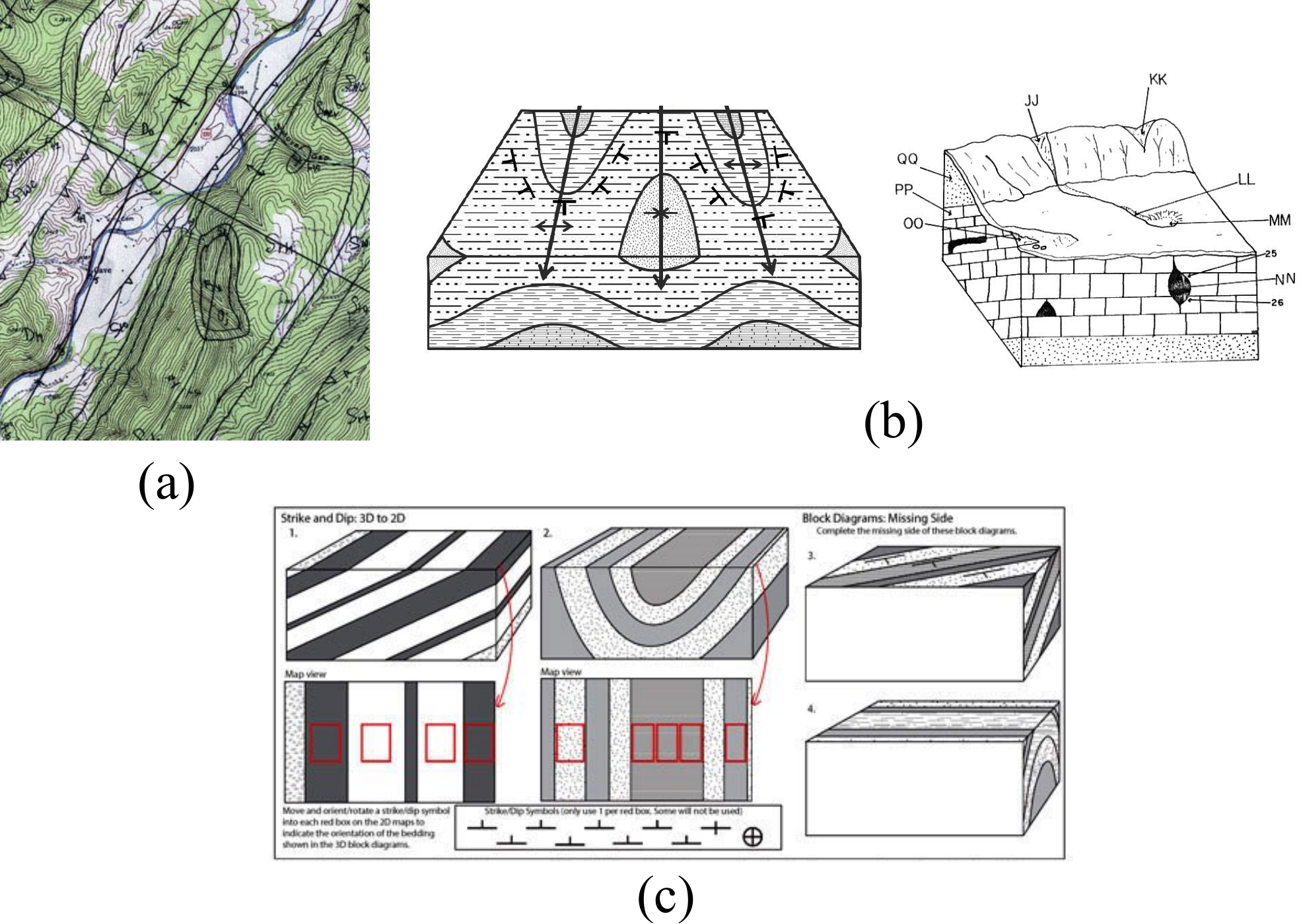}
    \caption{Sketching on geological maps is used extensively for spatial skills development, structural geology studies, data interpretation, and field exercises. 
    (a) Hand drawn geological features on USGS 1:24,000-scale topographic map \protect\cite{kurtDonaldson2003}.
    Source: West Virginia Geological and Economic Survey, Morgantown, WV; 
    (b) (left) Sketches reviewing contacts, strike, dip, and plunge; 
    (right) geological structure and topography in a sketched block diagram. 
    Source: Dr. Miriam Helen Hill, Physical and Earth Sciences, Jacksonville State University, Jacksonville, AL, 1990 and 2014; 
    (c) Worksheet covering structural geology concepts. 
    Students move and rotate the strike and dip symbols into the red boxes placed on top of geological maps to indicate the orientation of rock layers shown in the block diagrams. 
    For the Block Diagram section, students sketch the missing side of the diagrams by using the information on the other two sides, aiming at improving spatial skills perception \protect\cite{garnier2014}.
    Figure reproduced from Amorim et al.~\protect\cite{amorim2014}.}
     \label{fig:sketches}  
\end{figure}

Existing geological modeling tools are designed to allow the construction of very detailed models.
However, this increases the complexity of the tools, and need of specialized training.
Moreover, creating geological models is an interpretive task since the available data is limited and has a high degree of uncertainty.
The model construction workflow is usually divided in two stages: interpretation, and 3D modeling~\cite{jeanclaudeDulac2011}.
In the interpretation stage, skilled geologists fill in the gaps in data, using their knowledge and experience, by creating hand-drawn 2D sketches of their interpretation (Fig.~\ref{fig:sketches}).
Then 3D modeling experts create the actual geological model based on the available data and on the geologists' interpretation sketches.
After several iterations between these two stages the final model is constructed.

Creating geological models is a challenging task, and even simple models can take hours to be constructed.
The first challenge is the complexity of the models' geometry and topology.
Another challenge is the interpretative nature of the geological modeling task.
Data available, which may include for instance outcrops, well-logs, and 2D and 3D seismic, can be sparse and/or noisy creating uncertainty and making it prone to multiple interpretations.
Therefore, being able to try different interpretations and scenarios is an important step towards finding better approximations of the geology. 
However, usually only a single geological model is created due to the complexity and cost of the task~\cite{sEvans2003}. 

A different challenge in constructing geological models is the correctness of the model.
To ensure a model is correct from a geological perspective it should follow a set of geological rules.
For instance, \emph{horizons}, which are surfaces representing interfaces between two rock layers, should not contain self-intersections~\cite{caumon2009}.
Geologic models that do not comply with such rules can lead to serious and difficult-to-find problems in the final 3D models depicting the static (i.e. geology) and dynamic (i.e. fluid flow) mechanisms of reservoirs.

The modeling workflow itself is yet another challenge.
Because of the separation between interpretation and modeling tasks, during the interpretation stage, experts usually have to rely on conceptual, mental pictures of the 3D geology they are trying to interpret~\cite{caumon2009}.  
As the complexity of the geology increases, it becomes more difficult to make mental and conceptual models of the geology and many mistakes can be made.
These mistakes are passed on to the modeling stage where they are easily spotted on the actual 3D model, thus, increasing the number of iterations necessary to reach the final geological model.

The goal of this work is to mitigate the previously described challenges.
More specifically, we propose to narrow the gap between interpretation and modeling tasks by providing a system that enables the creation of 3D geological models directly from the interpretation sketches.
Therefore, the proposed system should enable trying different geological scenarios in a rapid modeling environment.
Moreover, it should create valid geological models by enforcing geological rules.
Finally, the system should be simple enough to not require additional training for the geologist.


To achieve the proposed goal, we propose a new sketch-based interface and modeling (SBIM) tool specifically designed for the rapid creation of geological models.
The basic idea of SBIM is to try to mimic the natural way people communicate ideas through 2D drawings and directly translating them into 3D models.
One of the main problems when creating a SBIM system is how to solve the ambiguity of going from a 2D sketch to a 3D model.
In general, each SBIM system defines its own \emph{sketching-language} specifying the set of sketches and annotations and how they define 3D shapes.
Although SBIM systems are meant to be straightforward, learning a new sketching-language still requires some effort.
Moreover, creating a specific sketching-language for geological modeling can be a complex task due to the complexity of the structures being modeled.
%

To deal with the problem of rapidly creating geological models from the geologists' 2D sketches we propose to use geological and topographic maps.
Geological and topographic maps are used by geologists to represent the 3D geological structures in a 2D media.
They have been used for many years and are part of the basic knowledge of any geologist.
Topographic maps are composed of contour curves defining the height of the terrain.
Geological maps are comprised of a 2D top view containing lines representing faults and contacts between rock layers, and standard geological symbols representing the 3D shape of these rocks.
A trained geologist can look these maps and understand the 3D geology in the subsurface.
In a sense, they are basically a sketching-language providing mappings from 2D lines and symbols into 3D geology.

In summary, this article proposes a SBIM system called Geo-Sketcher for creating 3D geological models directly from 2D sketches of geological and topographic maps.
Our main contribution is the new approach for rapidly creating 3D geological models from sketches of these types of maps.
Moreover, we embed geological rules in the sketching and 3D model generation in order to ensure validness of the model.
We also introduce the use of Hermite-Birkhoff Radial Basis Functions (HBRBF) for SBIM modeling.
Finally, we present a sketch-based approach for creating terrains on top of the geology based on topographic maps.

In \cite{amorim2014}, Amorim et al. introduce a sketch-based technique for 3D geological modeling.
They use concurrently sketch and annotations, commonly used by geologists, for rapid creation of 3D geological models.
In this article, we extend Amorim et al. work \cite{amorim2014} by including sketch-based modeling of the terrain from topographic maps sketches.
We also provide a detailed description of the techniques developed and used to create the 3D geological model from the sketch of geological and topographical maps, in special, the HBRBF interpolation applied to SBIM and geological modeling.

This article is organized as follows.
Related work is presented in Sec.~\ref{sec:related_work}.
Sec.~\ref{sec:background} provides background material in geological concepts including topographic and geological mapping.
Sec.~\ref{sec:system-overview} presents an overview of the Geo-Sketcher system.
Sec.~\ref{sec:line-pre-processing} describes how sketches are processed, while Sec.~\ref{sec:topographic-map-sketch}~and \ref{sec:geologic-map-sketch} present the methods developed for handling the sketch of topographic and geological maps.
Details of the construction of the 3D geological model from the 2D sketches are presented in Sec.~\ref{sec:creating-3d-geological-model}.
Sec.~\ref{sec:results_and_discussions} presents and discusses some results obtained using the proposed approach.
Finally, Sec.~\ref{sec:conclusions-and-future-work} presents the conclusions and future work.


\section{Related Work}
\label{sec:related_work}

The creation of geological models is discussed by Dulac in \cite{jeanclaudeDulac2011}.
He describes the workflow for creating these models as two separate tasks: interpretation and modeling.
The modeling task is dependent on the interpretation task and, as pointed out by Dulac, interpreters and modelers often disagree on each other's work.
The interpretation task is described by Bond et al. \cite{bond2007} as inherently uncertain, where different experts may produce different interpretation of the same geological structure.
Evans \cite{sEvans2003} advocates for a better integration between the interpretation and modeling tasks and\cite{sEvans2003} calls this gap in the modeling process of ``Valley of Death''.

Turner~\cite{turner2006} discusses the importance of recent advances on 3D modeling technologies, as well as requirements of geological modeling.
In his work, geo-objects are described as 3D objects of complex geometry and topology, with scale dependency and hierarchical relationships, and also having heterogeneous properties.
Therefore, specific geological modeling systems are required, since conventional modeling systems are designed to construct man-made structures with more regular shapes.
Moreover, geological structures configurations present specific rules that should be addressed by modeling systems to avoid geologically impossible scenarios, e.g., self-intersecting surfaces, as discussed by Caumon et al. \cite{caumon2009}, 

The use SBIM as a mean of providing more natural ways to create 3D models have been explored by different works~\cite{takeoIgarashi1999,cherlin2005,nealen2007,gunayOrbay2012}.
Olsen et al. \cite{lukeOlsen2011} use image-assisted SBIM of 3D objects combined with sketch-annotations. 
Modeling operations on the model being sketched are indicated by these annotations represented by iconic symbols.
Similar to our work, their system combines sketches and annotations for creating 3D models. 
Nevertheless, we propose to mimic how geologists sketch using geological and topographic maps instead of creating our own sketching language.
Besides, our system needs to enforce geological rules to avoid impossible geological scenarios.
Furthermore, geological symbols are composed of different parts, conveying different effects on the 3D model depending on their positioning, orientation, and size.

The application of SBIM to geological modeling has also been previously studied.
A SBIM system for modeling geological horizons is presented by Amorim et al.~\cite{amorim2012}.
The authors present a system with a set of sketch-based operators to model, modify and augment geological horizons.
Nevertheless, their focus is mainly on the augmentation of previously extracted horizons.
The construction of new horizons from scratch is limited to the creation of an initial Coons Surface. 
This surface usually needs a careful and long process of augmentation to get to the final desired shape.
The work of Lidal et al.~\cite{lidal2012} presents a comparison between two rapid sketch-based 3D geological modeling tools.
The tools presented are based on predefined features such as rivers and ridges/mountains or valleys and are based on the interpolation of user drawn curves on cross-sections and top view.

Other works have used sketch-based approaches to create simple illustration of geological scenarios.
Natali et al.~\cite{natali2012} use sketches of cross-sections to create rapid illustration of geology.
In  their work, the drawing is completely performed in a single cross section and no other information such as symbols are provided.
Thus, the generated 3D model is actually 2.5D, representing the extrusion of the cross-section.
Lidal et al.~\cite{lidal2013} present a system to communicate geologists' interpretation of seismic/slices (cross-sections) and how their interpretation was derived.
The authors present a flip-over metaphor system for sketching individual steps in a story reflecting the steps of the interpreter.
However, their work is intended to record the interpretation steps and not to create a 3D geological model.

For a further reading in geological modeling Natali et al.~\cite{natali2013} present a state-of-the-art report on terrain and geological modeling.
They compare types of surface representation and approaches to construct 3D geological models, including sketch-based modeling approaches.

\begin{figure*}[ht!]
    \centering
    \includegraphics[width=\linewidth]{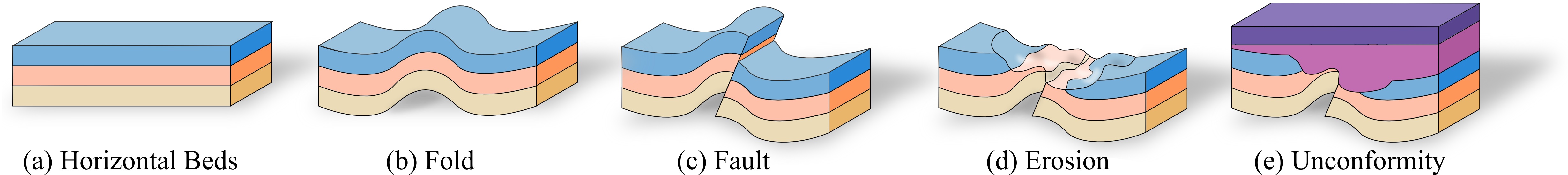}
    \caption{Illustration of geological concepts for the creation and deformation of rock layers.
    Layers are originally laid down horizontally through sediment deposition, then earth movements create folds and faults, and erosion removes part of the layers.
    Finally, sediment deposition restarts creating unconformities.}
    \label{fig:background}  
\end{figure*}
Finally, SBIM has also been used for creating terrains.
However, most systems are focused on the creation of visually realistic terrains for computer games, film special effects, training, and simulation.
Gain et al. \cite{gain2009} present a terrain modeling system that mixes interactivity with procedural methods.
The mountains and valleys are defined by silhouette, shadow, and region sketches.
Then the input silhouette sketch defining the mountain or valley has its noise characteristic propagated to the surrounding terrain generating a realistic heterogeneous terrain.
Hnaidi et al. \cite{hnaidi2010} present an approach based on user-defined feature curves with properties that are diffused to the rest of the terrain.
First a coarse terrain is constructed by feature curves that are able to define ridges, river beds, and cliffs.
Such constraints are then defined for the entire terrain domain by diffusion.
When the coarse terrain is ready, details are added by a procedural noise generator.
Each of theses works introduces its own sketching language which still needs to be learned.
Moreover, to the entertainment industry, terrain modeling techniques are focused on creating visually realistic models rather than creating geologically accurate terrains.

\section{Background}
\label{sec:background}

Geological models depict rock layers that have been deformed by earth movement.
Sedimentary rock layers, which are the focus of this work, are formed by the accumulation of sediments through a period of time.
When the type of sediment changes, a new layer is formed (see Fig.~\ref{fig:background}(a)). 
Following the law of superposition and principle of horizontality, these layers are laid down on top of each other forming relatively flat layers.
These layers are said to belong to the same \emph{rock series} and are transformed into rock by \textit{compaction} and \textit{cementation} (chemical process) \cite{alexMaltman1990,tarbuck2011}.
Sedimentary rocks are of particular interest because the process of piling organic materials with rock materials combined with appropriate pressure and heat through millions of years generates hydrocarbons such as petroleum \cite{tarbuck2011}, which are of great economic interest.

When these rock layers are subjected to tensional and compressional forces they can fold or fracture.
Folds are wave like deformations of the rock as depicted in Fig.~\ref{fig:background}(b).
And faults happen when the rock is fractured and displaced instead of folding (Fig.~\ref{fig:background}(c)).
Rock layers can also be shaped by erosion, which is caused by other agents such as water flow and wind.
Fig.~\ref{fig:background}(d) illustrates the rock layers deformation by erosion.

When sedimentation resumes after the existing rock layers are deformed, a new rock series is formed following the same law of superposition and principle of horizontality.
This new rock series is separated by the previous one by an \emph{unconformity} surface (Fig.~\ref{fig:background}(e)).


In this work we propose the rapid prototyping of 3D geological models directly from the geologists' 2D sketches.
More specifically, we aim at direct and interactive creation of 3D geological models by sketching geological and topographic maps.
Topographic maps represent the relief of the regions, while geological maps represent the geological structures of the surface and subsurface. 
These maps are aerial views of the area being represented and use standardized lines, symbols and colors to convey the desired information.
Topographic and geological maps can exist independently but are usually used in conjunction where the topographic map is the base map.
These types of maps are chosen because they offer a standard and well known approach for representing 3D geology in a 2D media.
In fact, being able to read and create such maps is part of the basic formation of geologists.
By using topographic and geological maps we minimize the problem of creating our own sketching-language, as well as the need of special training for using our system.
The next sections describe in further details these types of maps.

\subsection{Topographic Maps}
\label{sec:topographic-maps}

\begin{figure}[t!]
  \centering
  \includegraphics[width=\linewidth]{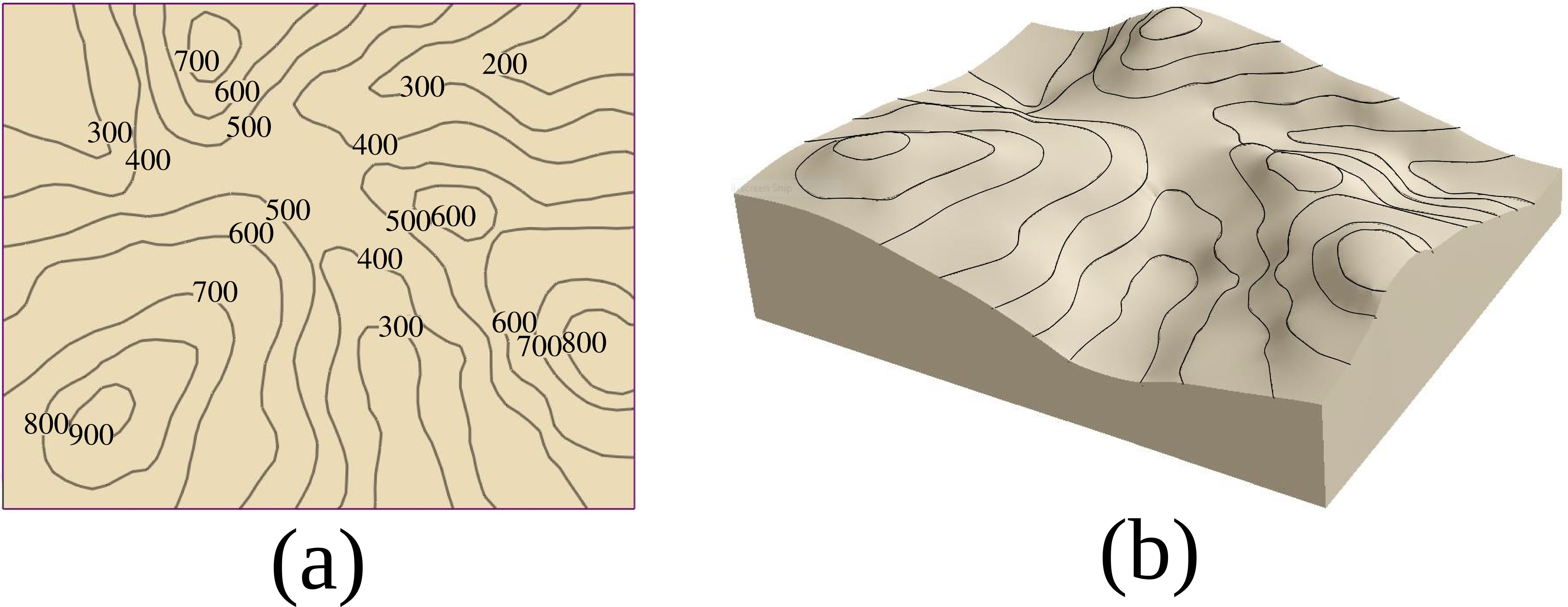}
  \caption{Example of a sketched topographic map in (a) and the resulting 3D terrain constructed in (b). This example was created using Geo-Sketcher.}
  \label{fig:topographic-example}  
\end{figure}

In this article we introduce sketch-based modeling of terrains using topographic maps, enabling the construction of more detailed geological models.
A topograpraphic map provides a standardized representation of 3D terrains in a 2D media.
Fig~\ref{fig:topographic-example} shows an example of a topographic map and its respective 3D terrain.
Topographic maps are basically comprised of contour curves, where each contour curve represents a constant elevation, i.e., all points in this curve have the same height.
Contour curves inside another contour curve are assumed to be higher than the containing curve, and contour curves elevations are assumed to be evenly spaced by a fixed amount.
Finally, a basic rule for constructing topographic maps is that contour curves should not have intersections or self-intersect.


\subsection{Geological Maps}
\label{sec:geologic-maps}

\begin{figure}[ht!]
\centering
\includegraphics[width=\linewidth]{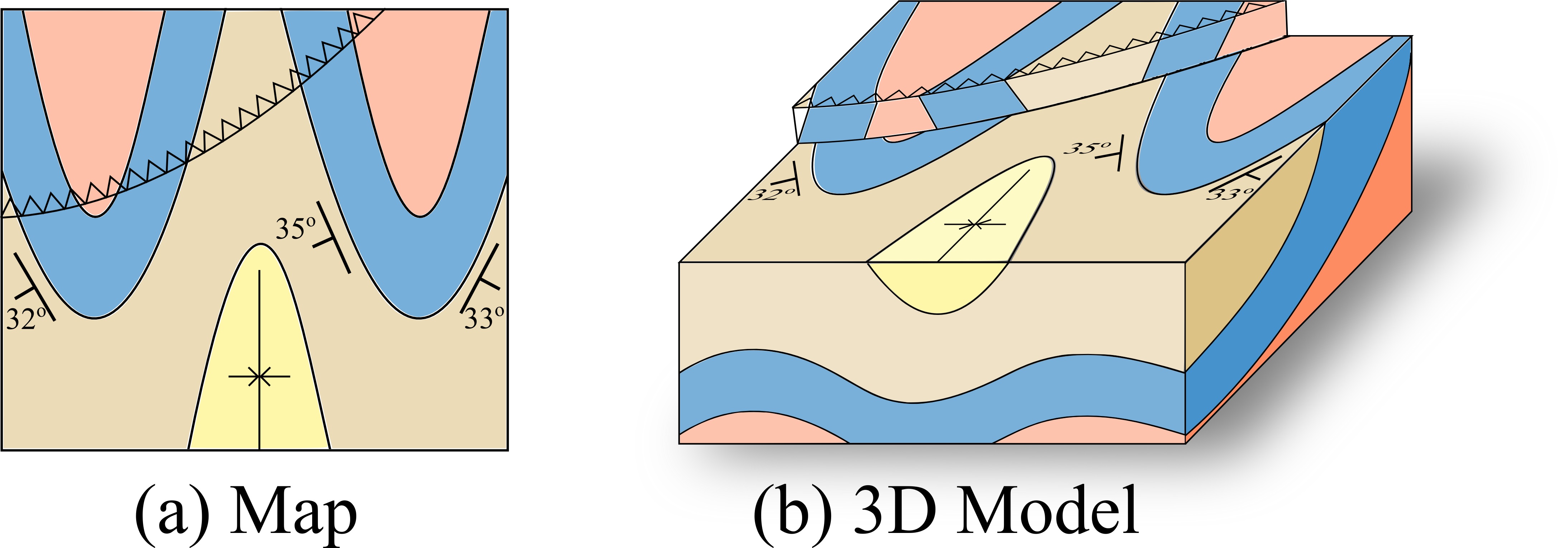}
\caption{Example of a geological map.
The map is composed of colors defining specific rock layers, lines separating these layers, and symbols defining the layers attitude.
In (a) the geological map, and (b) an illustration of the 3D model represented by the map.}
\label{fig:geological-map}  
\end{figure}

A geological map depicts the 3D geology through a 2D representation using lines, colors, and symbols.
This map is a top-view representation of the geology, where lines represent contacts between rock layers and rock series, colors represent different rock layers, and symbols define the deformation of the rock layers (see Fig.~\ref{fig:geological-map}).

Different geological structures such as dipping rock layers, folds, and faults are represented through geological symbols.
These symbols are used to indicate the structure's type, location and orientation~\cite{edgarSpencer1993}.
In the next sections we present examples of such structures in a geological map and what they represent in 3D.

\subsubsection{Dip and Strike}
\label{sec:dip-and-strike}

\begin{figure}[ht!]
    \centering
    \includegraphics[width=\linewidth]{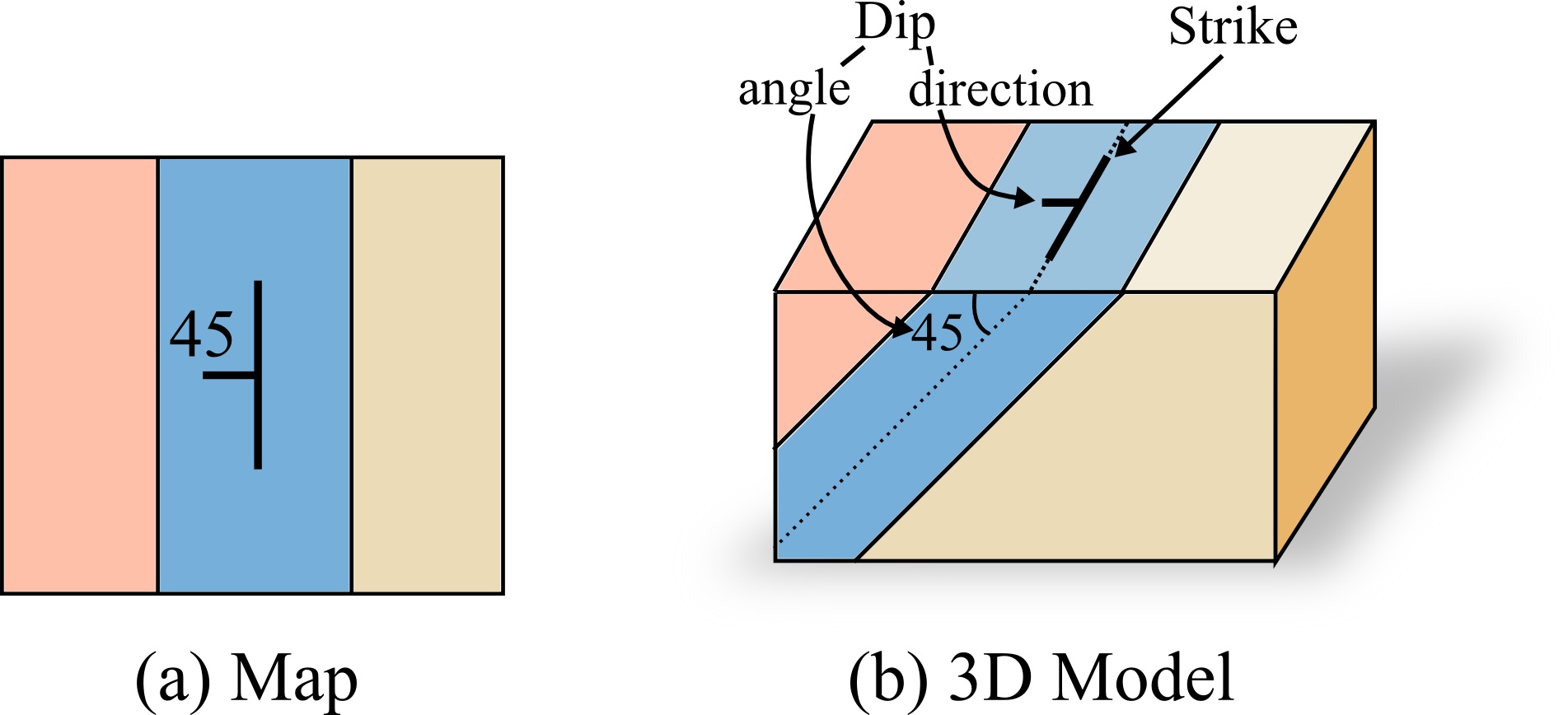}
    \caption{Dip and strike of an inclined rock layer. (a) a geological map with dip-and-strike symbol with $45^o$ dipping.
    (b) The resulting 3D model represented by the map and the dip-and-strike symbol parts explained.}
    \label{fig:dip_and_strike}  
\end{figure}

Rock layers are originally horizontal, however, as they are subjected to earth movements, they can tilt and change their attitude (see Fig.~\ref{fig:dip_and_strike}).
The dip of a rock is the angle that it makes with the horizontal plane and is between $0^o$ to $90^o$.
The dip direction is the direction in which such surfaces are inclined and can be visualized as the direction that a drop of water would follow if poured on the surface \cite{tarbuck2011,richardJLisle2004}.
The strike, in turn, is the direction of the line created by the intersection of the inclined surface with a horizontal plane.

2D geological maps present dip and strike by a symbol.
On a geological map the dip-and-strike symbol is represented by two lines and a number.
The number and shorter line specify the angle and direction of dip, while the longer line represents the strike (see Fig.~\ref{fig:dip_and_strike}~(a)). 
When geologists are on the field they try to measure as many as possible sedimentary rock layers dip and strike.
These data combined with description of rocks on a geological map or an aerial photograph can be used to infer the orientation and shape of the rocks~\cite{tarbuck2011}.

\subsubsection{Folds}
\label{sec:folds}

\begin{figure}[ht!]
  \centering
  \includegraphics[width=\linewidth]{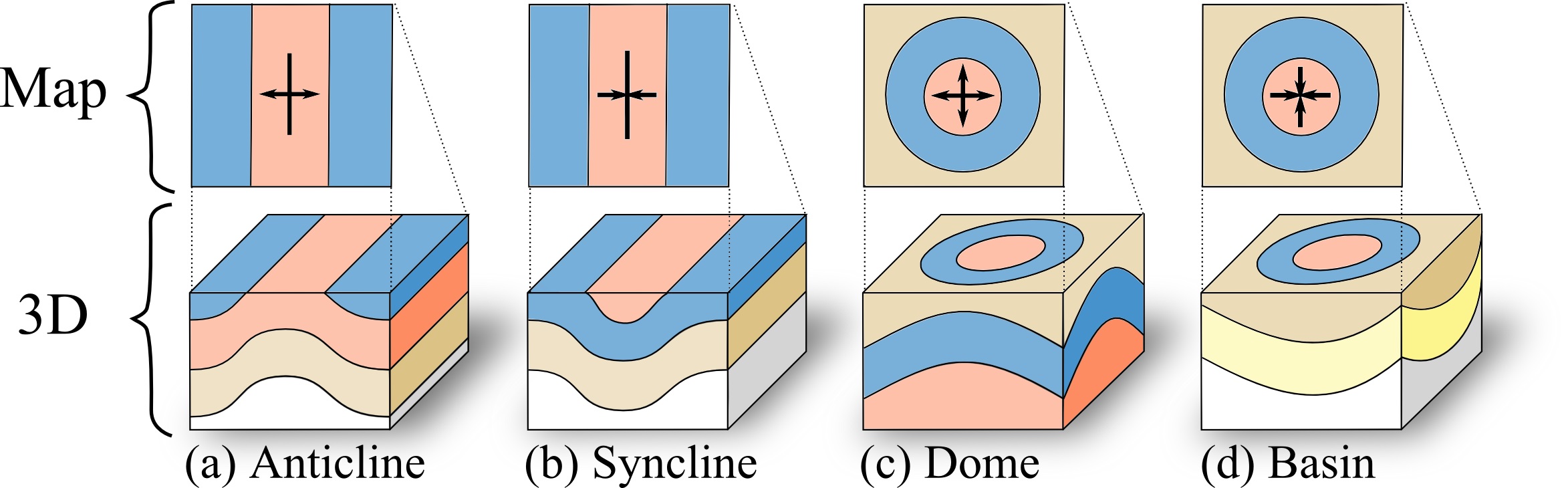}
  \caption{Different types of folds and their map representation.}
  \label{fig:folds}  
\end{figure}

Folds are geological structures that can be formed in virtually any rock type and depth (see Fig.~\ref{fig:folds}).
They are result of rock deformations caused by earth movements~\cite{alexMaltman1990,wejohnson1976}.
And can be of great economic interest in searching for minerals and hydrocarbons~\cite{haakonFossen2010}.

Folds are wave-like structures and, those cylindrical, can be classified as anticlines or synclines.
Anticlines are upfolds where older rocks are in the middle, and synclines are downfolds where younger rocks are in the middle (see Fig.~\ref{fig:folds}(a) and (b)).
Examples of folds that are not cylindrical include basins and domes.
Domes can work as hydrocarbon traps, and are have a shape similar to a cereal bowl turned upside-down \cite{haakonFossen2010}, i.e., the rock layers dip uniformly in all directions alway from the center.
Basins rock layers, on the other hand, dip in all directions towards the center.
When subjected to erosion, basins and domes create circular patterns on the surface (see Fig.~\ref{fig:folds}(c) and (d)).
Folds are represented in a geological map by symbols where anticlines and synclines have a variable length axis line accompanied of two arrows pointing outwards or inwards this axis, respectively (Fig.~\ref{fig:folds}(a) and (b)).
Domes and basins symbols are composed of four arrows all pointing outwards or inwards and do not have a variable length axis.

\subsubsection{Faults}
\label{sec:faults}

\begin{figure}[ht!]
  \centering
  \includegraphics[width=\linewidth]{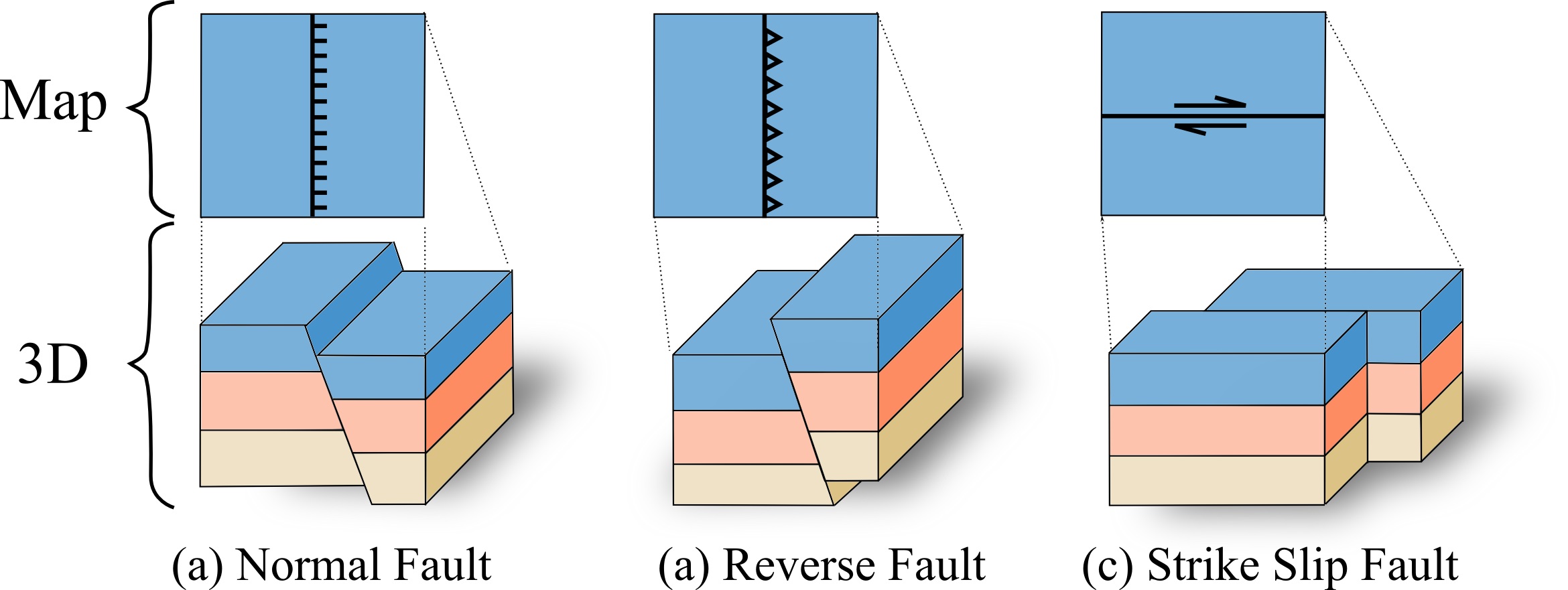}
  \caption{Three different types of faults and their map representations.
Normal and reverse faults are examples of dip-slip faults.
More specifically normal faults are caused by tensional stresses while reverse faults by compressional stresses.
Strike-slip faults are caused by shear stresses.}
  \label{fig:faults}  
\end{figure}

Rocks that undergo compressional and/or tensional forces can fracture instead of being continuously deformed by folding.
Faults are created when the rock fractures and there is a considerable displacement as result of earth movements~\cite{richardJLisle2004}.
Fig.~\ref{fig:faults} illustrates different types of faults and how they are represented in a geological map.

Common types of faults include dip-slip and strike-slip faults.
As suggested by their names, a fault category is defined by the slip direction of the rocks on the fault plane.
Dip-slip faults are caused by compressional or tensional forces displacing the rocks close to the vertical direction.
Strike-slip faults are those caused by shear stresses displacing the rocks close to the horizontal direction.
Fig.~\ref{fig:faults} presents two types of dip-slip faults and an example of strike slip fault.
Dip-slip faults can be further categorized as \textit{normal} or \textit{reverse} faults.
Forces pulling the rock in opposite directions are responsible for creating normal faults, while reverse faults are caused by compressional forces.
A fault's hanging wall is defined as the side of the fault above the fault plane.
Normal and reverse faults are represented in geological maps by geological symbols composed of a variable length strike line with triangles or dashes on the hanging wall side for normal and reverse faults respectively (see Fig.~\ref{fig:faults}).
In this work we implement the normal and reverse faults.

\subsection{Geologic Rules}
\label{sec:geological-rules}

When creating 3D geological models for simulations, specially fluid flow simulations of oil and gas, small problems in the 3D geological model structure directly affect the numerical model for the simulation, leading to completely wrong simulations.
Finding and repairing these problems later in the simulation stage can be a hard and time consuming task.
Many of these problems are caused by geologically invalid models.
Geological models are 3D solid objects; so, besides not having self-intersections, they should be water-tight, i.e., if represented by a mesh, this mesh should form a closed surface without any holes.
Moreover, each model is composed of different series of rock layers which are themselves 3D solid objects as well.
In order to form the geological model, the composing rock layers should fit perfectly on each other, i.e., where one rock layer ends another rock layer starts.

In this work, we propose to create valid geological models from the beginning, so that at any moment in the model construction, it is a valid model.
We tackle this problem in two separate parts of our modeling workflow:
(1) directly in the map sketch, preventing the expert to make mistakes; and, (2) by constructing the 3D model using methods that will guarantee that the resulting model is a solid.
The methods used to enforce the validness of the model in the 3D construction (2), are discussed in Section~\ref{sec:creating-3d-geological-model}.

\begin{figure}[ht!]
  \centering
  \includegraphics[width=0.7\linewidth]{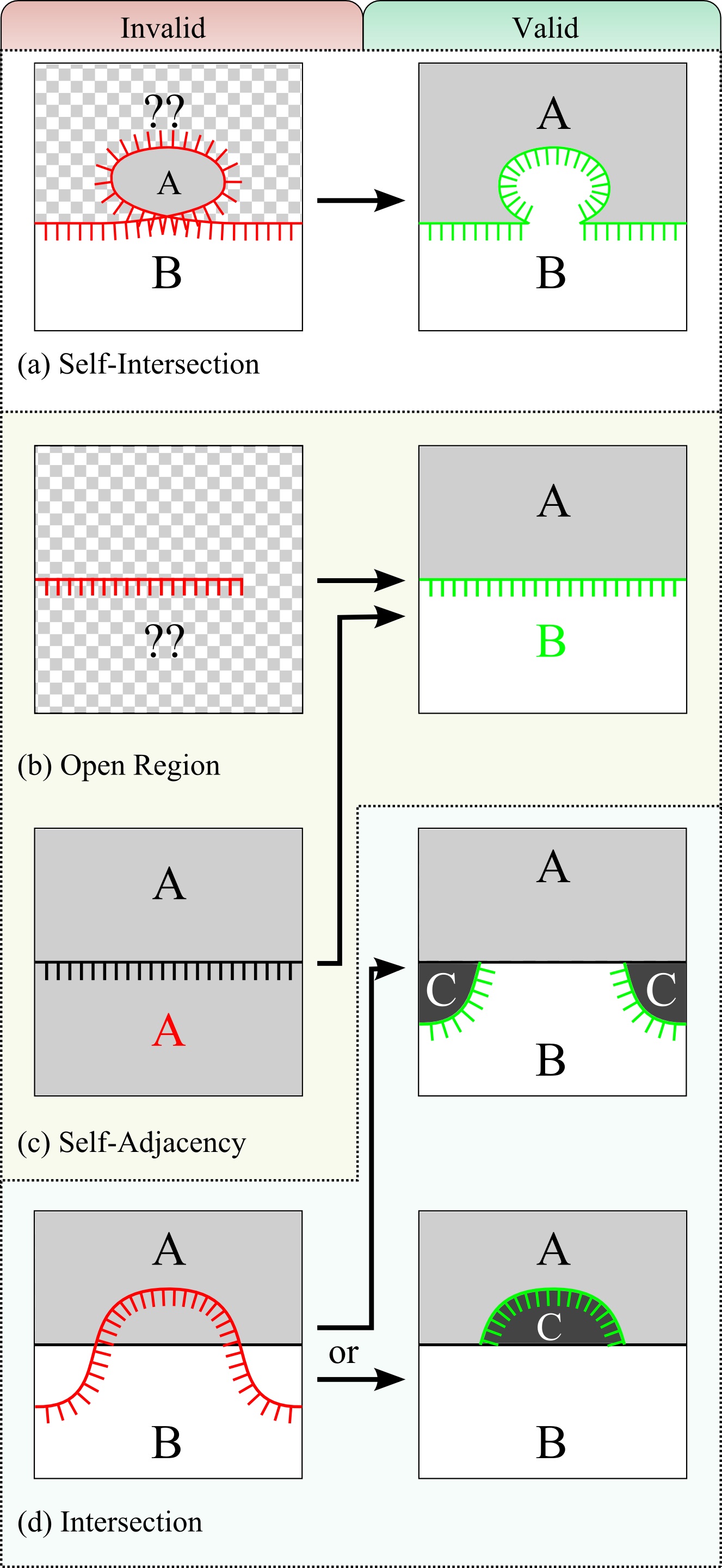}
  \caption{
  Illustration showing geological rules enforced directly in the geological map sketch. 
  Each contact line separates two and only two rock layers, and the hatches in the line show one side of the line.
  (a) self-intersection, creates a not well-defined side of the line.
  (b) open-region, there is no well-defined side of the line.
  (c) self-adjacency, a rock layer cannot be adjacent to itself.
  (d) intersection, hatches are ambiguous, new regions C is either on the hatched or the non-hatched sided of the line.
  }
  \label{fig:geologic-rules}  
\end{figure}

To apply geological rules directly in the geological map sketch, we enforce the four following rules:
(a) contacts must not self-intersect; (b) contacts always define closed regions on a map; (c) a rock layer cannot be adjacent to itself; (d) a contact must be interrupted after intersecting another.
Fig.~\ref{fig:geologic-rules} illustrates these rules showing examples of valid and invalid sketches.

For the topographic maps we enforce directly in the sketching the following three geological rules::
(a) contour curves must not self-intersect;
(b) contour curves must not intersect each other;
(c) contour curves inside other contour curves are higher.

\section{System Overview}
\label{sec:system-overview}
In few words, the Geo-Sketcher is a system that converts sketches of geological and topographic maps into a 3D geological model.
It is comprised of three main components: topographic map sketch, geological map sketch, and 3D model construction.
The input of the topographic and geological map sketch components are 2D sketches and their output is a set of interpolation constraints provided by the sketched contour curves, geological contacts, and symbols.
The 3D geological model component interpolates these constraints and creates the 3D geological model (see Sec.~\ref{sec:creating-3d-geological-model} for details).

\begin{figure}[ht!]
  \centering
  \includegraphics[width=\linewidth]{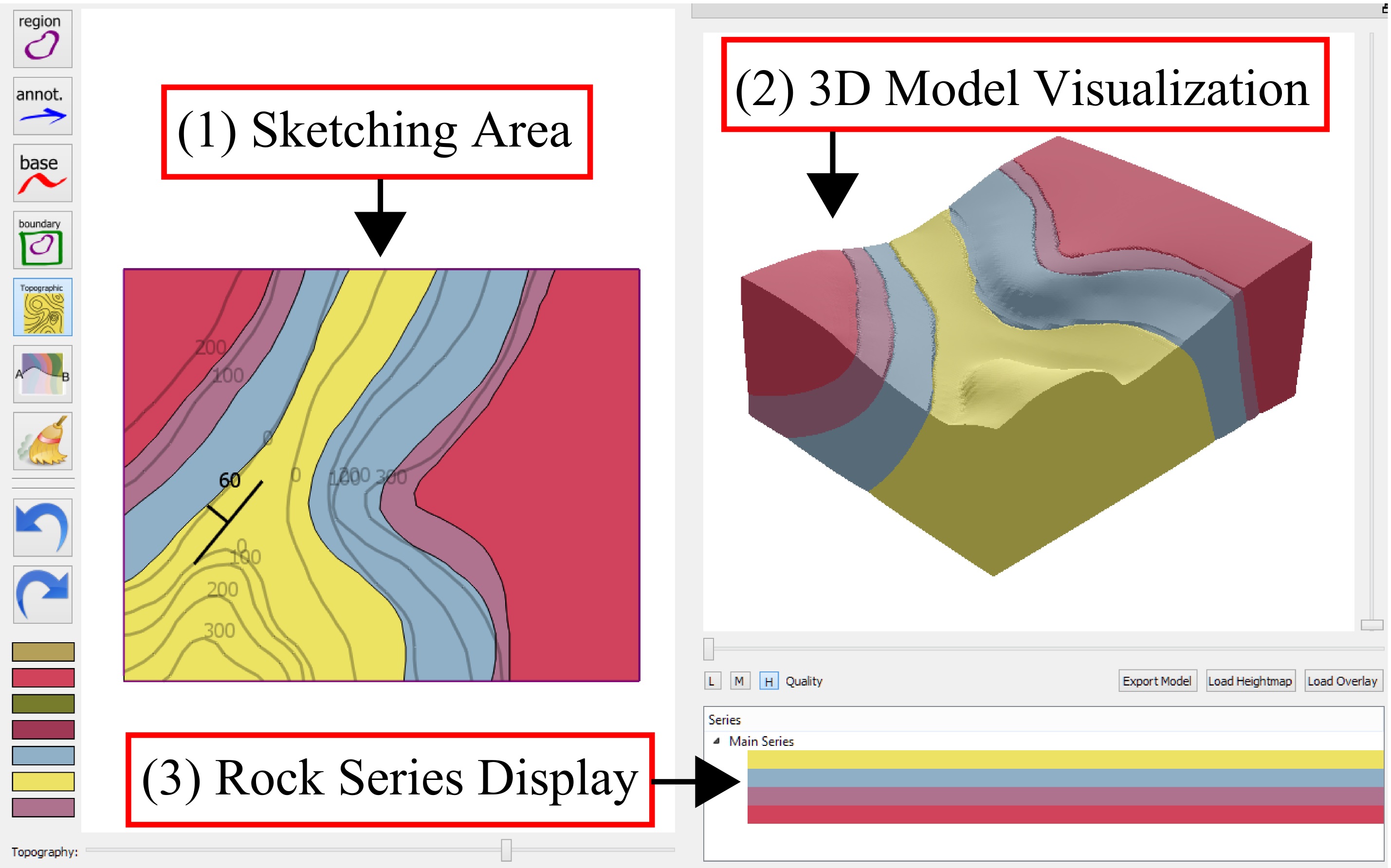}
  \caption{User interface is composed of 3 main windows and a tool-bar. 
  (1) where geological and topographic maps are sketched.
  (2) presents the 3D model created from the sketches.
  (3) displays the rock series automatically computed from the sketches.
  }
  \label{fig:interface}  
\end{figure}

The Geo-Sketcher interface is comprised of a few buttons, a sketching area, a rock series display, and the 3D model visualization (see Fig.~\ref{fig:interface}).
The buttons include different pens for sketching geological contacts, symbols, topographic contour curves, and cross-sections cuts, as well as tools for undo/redo and selecting the rock layer.
The sketching area allows the sketching of topographic and geological maps.
As the maps are sketched, the respective 3D model is created/updated and shown in the 3D model visualization window.
The rock series display window shows the automatically computed relationship between rock layers and series, where the layers and series on top are the oldest.

The next sections describe the methods used to handle the sketching of topographic and geological maps and how these sketches are translated into a 3D geological model.


\section{Line Pre-Processing}
\label{sec:line-pre-processing}

Sketching lines is the main input type of the Geo-Sketcher system. 
Each sketch is a line created by connecting consecutive non-uniformly spaced points provided by the input device, which could be either a pen, for pen-enabled computers, or a mouse.
In this work, to allow different levels of details for the lines, when a line is sketched it is pre-processed taking into account the zoom factor.
So that lines with more details can be sketched by zooming in the sketching area.
The sketch pre-processing step comprises of first super-sampling it by equidistant points based on the zoom factor, to take into account the level of details.
And then applying a reverse Chaikin filter~\cite{cherlin2005} four times to remove noise from the input.

This sketch pre-processing is applied for all lines sketched in Geo-Sketcher for either topographic or geological map sketches, as well as symbols.

\section{Topographic Map Sketch}
\label{sec:topographic-map-sketch}

This sketching mode enables the creation of 3D terrains using the geologist's standard notation of topographic maps.
When in topographic map sketch mode (by selecting the appropriate pen in the user-interface), all sketches are interpreted as contour curves. 
The next step is to start sketching the contour curves and associating a height value for each.
One approach would be to request the user to define the height for each curve as they are drawn.
However, this can be a time consuming task and lead to violations of the topographic maps rules, where contour curves should be evenly spaced vertically.
The other approach, which we propose and use in this work, is to request only the contour curve height displacement and the first contour curve height.
All other contour curves can then be computed based on this information.
So, when the expert draws the first contour curve of the topographic map, the system requests the height of this contour curve and the displacement between contour curves to be hand-written.
The hand-written digits are recognized using the approach presented in the end of Section~\ref{sec:symbol-recognition}.
\begin{figure}[ht!]
  \centering
  \includegraphics[width=\linewidth]{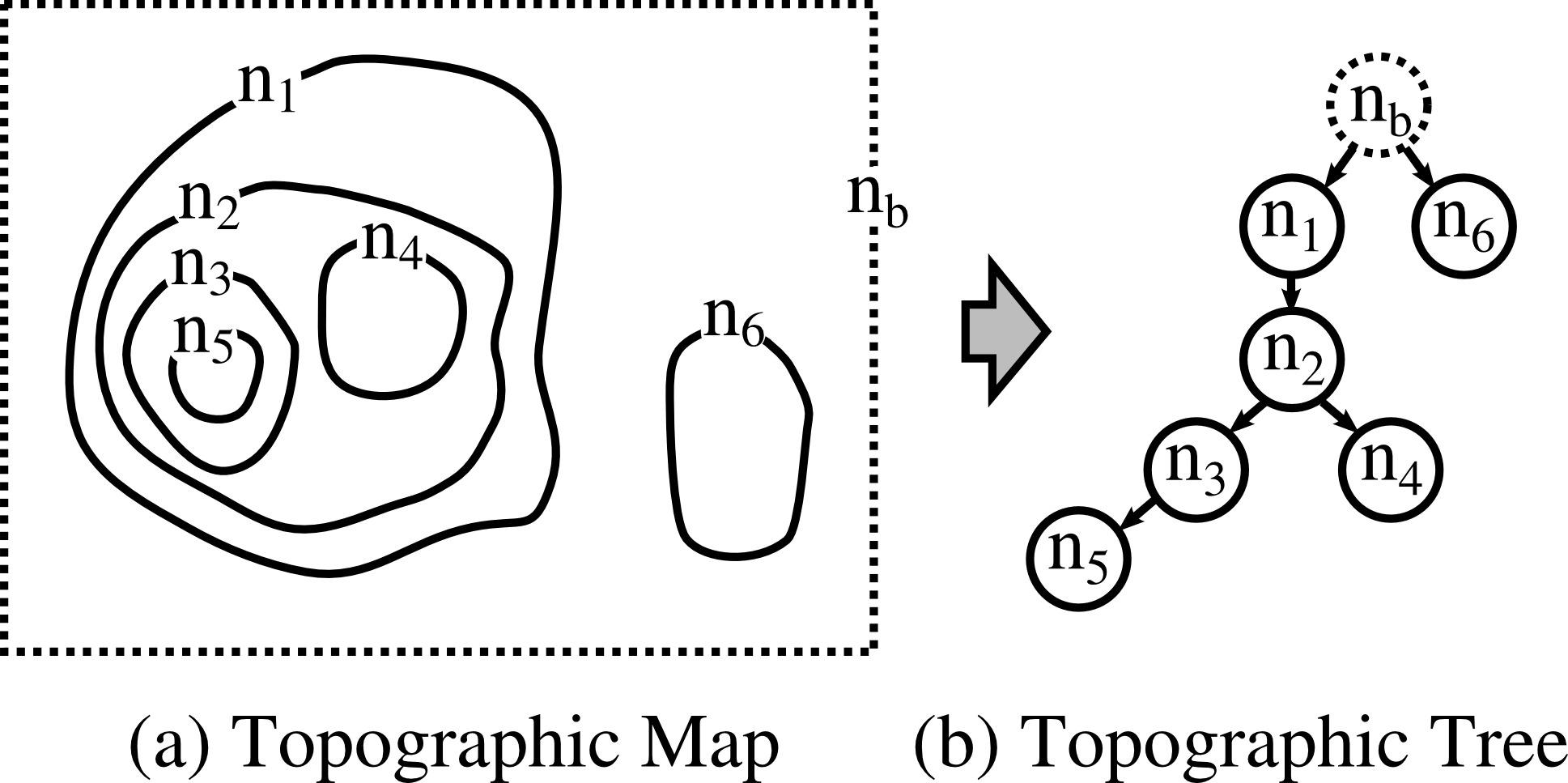}
  \caption{Representation of a topographic map in (a) and its respective topographic tree.}
  \label{fig:topographic-tree}  
\end{figure}

We developed an algorithm that stores the contour curves in a tree data-structure to compute the height of all contour curves based on the two values provided and the construction rules of topographic maps.
Since contour curves must not intersect, each contour curve can be enclosed by one other curve and/or enclose other contour curves.
Let's call a contour curve $X$ \emph{parent} of another contour curve $Y$ if $X$ encloses $Y$ and no other contour curve enclosed by $X$ encloses $Y$.
In this case, $Y$ is called a \emph{child} of $X$.
To illustrate these concepts see Fig.~\ref{fig:topographic-tree}(a), where curve $n_2$ is the parent of $n_3$ and $n_4$, but $n_1$ is only parent of $n_2$.
Based on this parent/child definition we construct a tree as shown in Fig.~\ref{fig:topographic-tree}(b).
Where the root node is always the boundary $n_b$ defining the sketching area.
This contour curve is composed of only points with no height associated (represented by the dashed lines in Fig.~\ref{fig:topographic-tree}(a)).
So each child node is enclosed by its parent node, therefore, each child-node's height is the parent-node's height plus the displacement.
Since the first contour curve height was defined, it should be fixed, so all others are computed based on it.
For instance, let's suppose that the node $n_3$ in Fig.~\ref{fig:topographic-tree} was the first contour curve defined and had its height defined as $400m$ with a displacement of $100m$.
In this case our algorithm would traverse the tree and compute the height of $n_2$ as $300m$, $n_4$ as $400m$, $n_5$ as $500m$, etc.
\begin{figure*}[ht!]
  \centering
  \includegraphics[width=\linewidth]{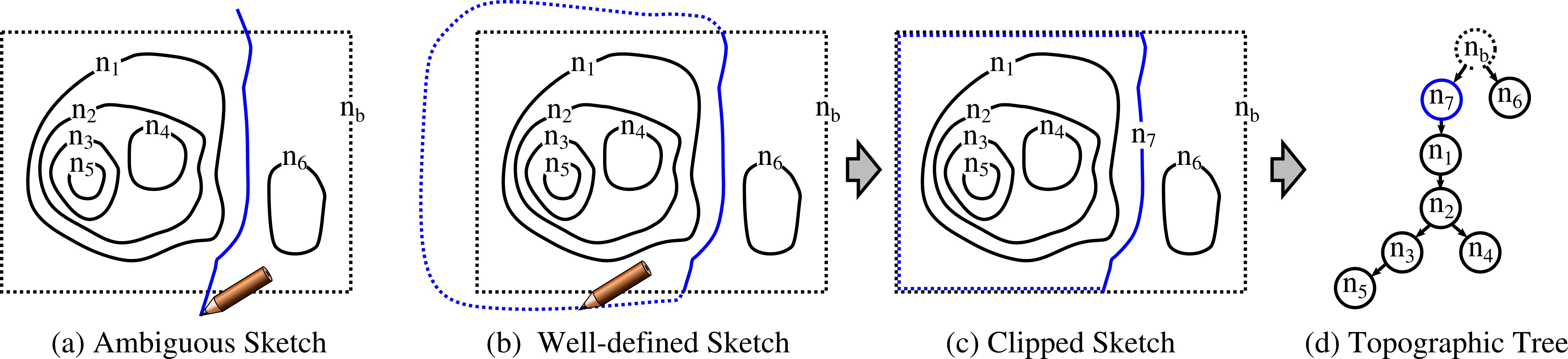}
  \caption{Ambiguous sketch in the topographic map.
  (b) The ambiguity is resolved by continuing the sketch enclosing the desired region.
  (c) The sketch is clipped to the boundary of the map.
  (d) The appropriate topographic tree is constructed.}
  \label{fig:topographic-ambiguity}  
\end{figure*}

The last thing to consider is the intersection between contour curves.
Contour curves must not intersect other contour curves or self-intersect.
The intersection between contour curves is handled as the curves are sketched so that the sketching only resumes when the sketching pen returns to a non-intersection point.
Clearly, using this approach only the new contour curve parent and the parent's children needs to be checked for intersection, improving the performance and enabling intersection check as the sketch is created.
The only type of intersection allowed are those with the boundary contour curve.
In which case, the intersecting contour curve is clipped to be within the boundary.
However, contour curves intersecting the boundary can lead to ambiguity problems when constructing the topographic tree.
Fig.~\ref{fig:topographic-ambiguity}(a) shows an example of ambiguity, where the new sketched contour curve $n_7$ can be a parent of $n_1$ or $n_5$.
To solve this ambiguity without requiring the definition of the height for the $n_7$ contour curve, the system requires the expert to define his/her intention by continuing the sketch outside boundary and enclosing the desired curve (see Fig.~\ref{fig:topographic-ambiguity}(b)).
The boundary contour curve clips the intersecting contour curve according to the experts intention, and the clipped curve is replaced by new dummy points as the points belonging to the boundary (see Fig.~\ref{fig:topographic-ambiguity}(c)).
This dummy points are necessary to clearly define the enclosing and to ensure the validness of the topographic map.

The creation of the 3D terrain is explained in Section~\ref{sec:rbf}, but the basic idea is to find a surface that interpolates the contour curves of the topographic map.
However, instead of interpolating the curves, we interpolate a set of points along the contour curves, which are generated using the line pre-processing described in Section~\ref{sec:line-pre-processing}.
Each of these points provides information about its $(x, y, z)$ position where $(x, y)$ are the coordinates on the topographic map and $z$ is the height of the contour curve.

\section{Geological Map Sketch}
\label{sec:geologic-map-sketch}

\begin{figure}[ht!]
  \centering
  \includegraphics[width=0.8\linewidth]{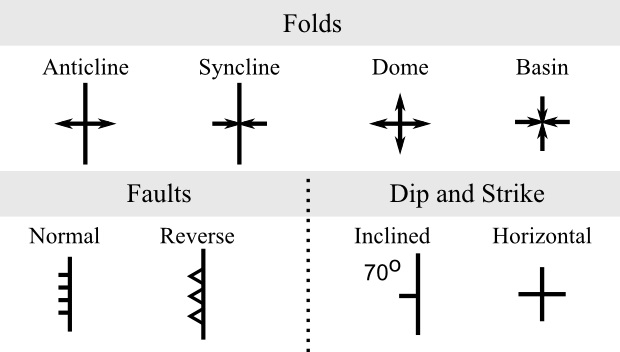}
  \caption{Symbols currently supported by Geo-Sketcher.
  The meaning of each symbol is described in Section~\ref{sec:background}.
  The horizontal rock layers symbol represents layers with no dipping.}
  \label{fig:symbols}  
\end{figure}

The geological map sketch can be divided into contacts and symbols sketch.
The interface provides pens that enable the sketching of geological contacts and/or geological symbols.
In order to construct a valid 3D geological model and a valid geological map, we need to enforce geological rules, described in Section~\ref{sec:geological-rules}, directly in the sketching.
We accomplish this task by creating a contacts-graph (see Sec.~\ref{sec:contacts-graph}) that ensures the model is correct.
Finally, to create the 3D model we need to recognize the hand-drawn geological symbols.
%
Fig.~\ref{fig:symbols} presents a set of the geological symbols supported in Geo-Sketcher, which are: dip-and-strike, horizontal rock layers, anticline, syncline, dome, basin, normal-fault and reverse-fault.
However, other symbols can be included using the same approach proposed here.

The layers of the 3D geological model being constructed are defined by surfaces that interpolate the sketched contacts after they have been projected on the terrain defined by the topographic map.
The attitude and shape of these surfaces is dictated by the interpolation of the extra information provided by the geological symbols.
The details of how the contacts-graph and geological symbols define these interpolation constraints and how they are used to create the 3D geological model is presented in Sec.~\ref{sec:creating-3d-geological-model}.
The next sections will describe how contacts-graphs are created and how geological symbols are recognized.

\subsection{Contacts-Graph}
\label{sec:contacts-graph}
\begin{figure}[ht!]
  \centering
  \includegraphics[width=\linewidth]{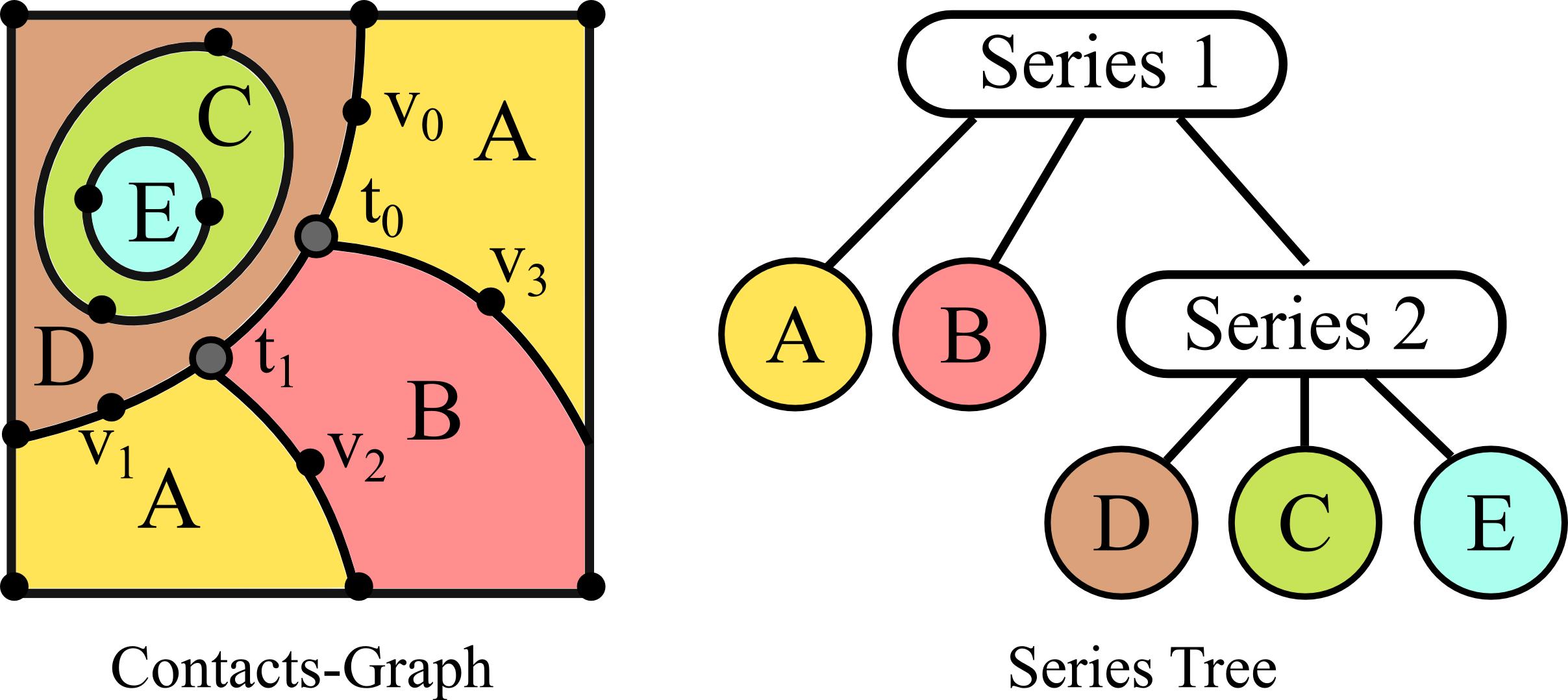}
  \caption{Illustration of contacts-graph and corresponding \emph{series-tree}. In the contacts-graph, vertices $t_0$ and $t_1$ lists of neighbors are both $\{A, B, D\}$.
  vertices $v_0$ and $v_1$ lists of neighbors are both $\{A, D\}$; vertices $v_2$ and $v_3$ lists of neighbors are both $\{A,B\}$;
   the T junctions $t_0$ and $t_1$ indicate that the contact containing $v_0$ and $v_1$ is defined as an unconformity since.}   
  \label{fig:graph-and-tree}  
\end{figure}

To capture the sketch of geological contacts, we developed a graph-based representation which is used to solve three different problems.
The first problem is checking the compliance with geological rules (see Sec.~\ref{sec:enforcing-geological-rules}).
The second problem is to automatically find the order of the rock layers, and define different rock series by detecting unconformities (see Sec.~\ref{sec:automatically-finding-rock-layers sequence}).
The third is recording the sketched rock contacts and providing interpolation constraints for constructing the 3D geological model (see Sec.~\ref{sec:creating-3d-geological-model}).
In our undirected graph representation each vertex contains information of its 2D position on the map, a list of neighboring rock layers, and information whether it belongs to a contact or to the boundary of the map.
An example of a contacts-graph is shown in Fig.~\ref{fig:graph-and-tree}.

The contacts-graph is first initialized with a non-self-intersecting closed curve, which is usually a rectangular shape, representing the boundary of the geological map.
This boundary defines the limits of the map and is necessary to enforce the geological rule which requires sketched contacts to define closed regions.

\subsubsection{Enforcing Geological Rules}
\label{sec:enforcing-geological-rules}

After a new contact is sketched, it is first pre-processed, as described in Section~\ref{sec:line-pre-processing}.
The next step, is to check whether the sketched contact is compliant with geological rules (described in Sec.~\ref{sec:geological-rules} and restated here for convenience):
(a) contacts must not self-intersect; (b) contacts always define closed regions on a map; (c) a rock layer cannot be adjacent to itself; (d) a contact must be interrupted after intersecting another.
At this point the line sketched is checked for self-intersections to make sure it is not violating geological rule~(a).
After checking problems with the sketched line, we need to verify whether the sketch violates geological rules when added to the geological map.
A new sketch can fall in five cases when trying to be added to the contacts-graph:
\begin{enumerate}
\item [(1)] It is totally outside the boundary of the contacts-graph, which is not allowed.
\item [(2)] It has one intersection with the contacts-graph, which violates geological rule~(b).
\item [(3)] It has two intersections with the contacts-graph, which is allowed.
\item [(4)] It has more than two intersections which violates rule~(d), so it needs to be trimmed to only two intersections.
\item [(5)] It is completely inside the contacts-graph and it does not have any intersection, the sketch will be closed in a loop to comply with rule~(b).
\end{enumerate}

If the sketched curve passes the validation process it will define a new closed region representing a rock layer in one side and another rock layer on the other side.
If both sides are defining the same rock layer the sketched curve is not following geological rule~(c) and it needs to be corrected.
Finally, when all geological rules are respected, the contacts-graph is updated to include the sketched contact.

\subsubsection{Automatically Finding Rock Layers Hierarchy}
\label{sec:automatically-finding-rock-layers sequence}

In order to expedite the modeling process our system automatically computes the sequence of rock layers and series based on the provided sketches.
By analyzing the rock layer neighboring information provided by the vertices on the contacts-graph we can construct a linked list of layers.
To illustrate the idea consider the rock layers $C$, $D$, and $E$ shown in Fig.~\ref{fig:graph-and-tree} and the respective linked list $D \leftrightarrow C \leftrightarrow E$.
However, this linked list is ambiguous since it does not tell whether $E$ is the oldest or youngest layer.
To solve this ambiguity we choose one contact which is not defining an unconformity surface.
Then we create an approximated 3D surface  of the chosen contact using all symbols belonging to contact's rock series (see Sec.~\ref{sec:hbrbf}).
Finally, we select a point on the terrain inside the closed region defined by this contact and check whether it is below or above this 3D surface.
If the point is below this surface, $E$ is the oldest rock layer, otherwise it is the youngest.
The successful construction of this list is also used to check an additional geological rule where a rock layer can only be adjacent to a maximum of two rock layers in the same series.

Unconformities are also automatically detected using the contacts-graph information.
In order to decide whether a closed region is an unconformity we look for \emph{T} junctions vertices and check their adjacent vertices neighboring rock layers list.
Based on these unconformities we build the \emph{series-tree} composed of unconformities and rock layers (see Fig.~\ref{fig:graph-and-tree}).
This series-tree represents the hierarchy of unconformities (older on top) and all leaves are rock layers.
This tree is used to decide to which series a rock layer belongs and to properly construct the 3D geological model (see Sec.~\ref{sec:creating-3d-geological-model}).

\subsection{Symbol and Hand-Writing Recognition}
\label{sec:symbol-recognition}
There is a great variety of symbol recognizers aiming to single stroke symbols \cite{jacobWobbrock2007,herold2012,lukeOlsen2007,deanRubine1991}.
Geological symbols, however, are composed of different strokes that can be drawn in several different ways depending on the order and direction of the strokes.
The problem in multi-stroke symbol recognition is how to combine multiple strokes into a single symbol.
Some approaches simplify the problem by requiring the strokes to be drawn on a specific order~\cite{donmez2012}.
Others, solve this problem using a combinatorial approach~\cite{lisaAnthony2010,lisaAnthony2012}.
Image-based recognizers, on the other-hand, do not suffer the same problem with multi-stroke symbols~\cite{leventKara2005,heloiseHse2004}, but they cannot be directly applied to our problem because geological symbols may be composed of parts that vary in shape and length.
To the best of our knowledge, none of the existing multi-stroke symbol recognizers can be directly used to recognize geological symbols.

\begin{figure}[t!]
  \centering
  \includegraphics[width=0.4\linewidth]{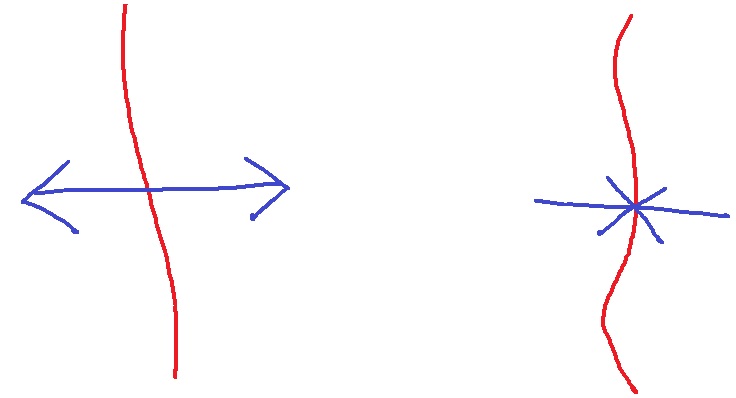}
  \caption{Sketch of geological symbols using two pens. The fixed part of the symbol is sketched with the blue pen while the varying one (strike stroke) is sketched with the red pen.} 
  \label{fig:geological_symbols}  
\end{figure}

To mitigate this problem of recognizing a symbol that has a part that varies in shape and length, we separate the symbol sketching in two parts.
A first pen is used to sketch the varying part of the symbol, and a second pen is used for the fixed part.
Fig.~\ref{fig:geological_symbols} presents examples of geological symbols that are recognized with the varying part in red, which we call strike stroke, and the fixed part in blue.
For the symbols considered in this work, the varying part is always a single curve of any length and shape.
The fixed part, on the other hand, can be composed of multi-strokes. 
To recognize the geological symbols we developed a multi-stroke symbol recognizer that combines features coming from the relationship between the varying part with the fixed part, as well as features extracted exclusively from the fixed part of the symbol.

Our multi-stroke symbol recognizer is based on the approach proposed by Glucksmann~\cite{glucksman1967}.
In our approach each stroke is defined by a line, pre-processed using the method described in Section~\ref{sec:line-pre-processing}, containing a sequence of 2D points that will be compared against a dictionary of template symbols.

To compare the input strokes against a template symbol, we compute the axis of maximum variance of all strokes' points using Principal Component Analysis (PCA).
The coordinates of the points are then rewritten using the PCA center and axes, making it translation and rotation invariant.
Then we scale these points to fit in the unit square making the symbol scale invariant.
Since the symbol orientation is important to our application, we extract this information from the PCA angle and later use it as part of the 3D model computation as well as in the drawing of the template symbol on the geological map canvas.
The next step is discretizing the $x$ axis (principal component) into 64 bins.
For each bin $b$ the number of intersections of the symbol strokes with the bin vertical line $l_b$ is computed, where $l_b = (\frac{2b + 1}{128}-0.5, y)$.
However, since we are interested in finding the intended angle that a symbol was drawn, the PCA alone can lead to problems.
As an example consider two dip-and-strike symbols, where the first  points up while the second points down. 
The vectors $(1,0)$ and $(0,1)$ are valid principal components for both symbols.
Therefore, if we use these same vectors to compute the angle of both symbols both would yield $0^\circ$, where we know the second is $180^\circ$.
In order to solve this ambiguity, we need to orient both vectors consistently.
We accomplish this by computing the density of points on left and right quadrants and orient the first vector to point in the direction of greater density.
The same is done for the second vector, but now based on the top and bottom quadrants.

Finally, two more features are important to compare symbols.
First, the width and height aspect ratio.
Second, the intersection with the strike stroke, which includes the number of intersections, as well as their distance compared to the template symbol.
In order to combine the three features we use the following formula:
\begin{equation}
\begin{split}
 D_t = &\underbrace{w_v \lVert \vec{v_t} - \vec{v} \rVert^2}_\text{vertical intersections} + \underbrace{w_r \lvert r_t - r\rvert^2}_\text{width/height ratio} \\
& + \underbrace{w_i (\lVert \vec{i_t} - \vec{i} \rVert^2 + \lvert n_t - n \rvert P_{n})}_\text{intersections},
\end{split}
\end{equation}
where the subscripts $t$ represent the feature of the template symbol,
$v$ represents the vertical intersections vector, $r$ the width to height ratio, $i$ the intersection point, $n$ the number of intersections, $P_n$ is a penalty for having a different number of intersections from the template symbol, and $w_v$, $w_r$, and $w_i$ are weights.
For the symbols we need to recognize we found experimentally that a good choice of parameters is: $w_v = 0.273$, $w_r = 0.273$, $w_i = 0.454$, $P_n = 3.0$.
The recognition threshold is also experimentally set to $21$, which means that if for all $t$ $D_t>21$ the sketched symbol will not be recognized.
We have a dictionary of template symbols that we want to match.
They are defined in our system in a text file that specifies lines of the fixed part of the symbol, whether or not the symbol has a strike-stroke and where the intersections are positioned.
Using this approach new symbols can be easily added to the interface in case we want to support more geological symbols.

Finally, symbols such as dip-and-strike and faults require the specification of numeric values.
These numeric values are prompted to be handwritten when these symbols are recognized.
To recognize handwritten values we rely on the Tesseract OCR engine~\cite{tesseract} trained with samples of handwritten digits.

\section{Constructing the 3D Geological Model}
\label{sec:creating-3d-geological-model}

In our system the 3D geological model is created by surfaces interpolating the data provided by the topographic and geological maps.
To construct this model we need to solve three main problems:
(1) create a surface, representing the terrain, which interpolates the constraints from the sketched topographic map (see Sec.~\ref{sec:rbf});
(2) create a set of surfaces, representing the interfaces between the rock layers, which interpolates the constraints from the sketched geological map (see Sec.~\ref{sec:hbrbf} and~\ref{sec:iso-value-estimation});
and (3) combine these surfaces, terrain and rock layers interfaces, to construct a water-tight 3D representation of the model (see Sec.~\ref{sec:faults-construction} and~\ref{sec:csg}).

In the surface interpolation problem, the sketched topographic map defines the geometry of the top of 3D geological model through contour curves (discretized into points), and serve as a base to where the geological map contacts and symbols are projected. 
These contacts (discretized into points) define where the interfaces between two rock layers are cutting the terrain and the symbols describe the attitude of the rock layers at the specified point.

For solving problem (2) we need an interpolation method that enables the interpolation of points in space (defined by contacts) and dipping information at a point in space. Where the dipping information affects the attitude of the surfaces interpolating the contacts.
Another constraint imposed by the nature of the object being modeled is that the surfaces representing the rock layers interfaces in a same rock series are assumed to be sub-parallel to each other. This requirement is given by the law of superposition and principle of horizontality (see Sec.~\ref{sec:background}).
To tackle these problems we propose the use of Hermite-Birkhoff Radial Basis Functions (HBRBFs) interpolation by Macedo~\cite{macedo2011}.
This type of implicit surface allows us to interpolate functions given three different types of constraints: function values at a point in space, function gradient at a point in space, and function value \& gradient at a point in space.
Details about the mathematical formulation of this interpolation method are presented in Sec.~\ref{sec:hbrbf}.

The mapping of the geological map constraints to the HBRBF formulation is mostly directly.
The HBRBF define a potential-field, which provides a good approximation of sub-parallel iso-surfaces.
And, since rock layers in a same rock series are assumed to be sub-parallel, we use a single HBRBF to interpolate each rock series.
The rock layers interfaces are defined as iso-surfaces of the HBRBF function which interpolates the respective contacts on the geological map.
Therefore, the contacts points constraints are mapped in the HBRBF interpolation as points with values.
Where these values are associated with the iso-surfaces of the rock layer interface.
More details of how to choose these iso-values are presented in Sec.~\ref{sec:iso-value-estimation}.
The dip-and-strike symbols provide information that can be easily transformed into points with gradient information and directly used as constraints for the HBRBF interpolation.
The other geological symbols representing interpretative fold symbols such as domes, basins, anticlines and synclines, have their strike curve mapped into points with values \& gradients.
However, the choice of value is made such that it distorts the potential-field up or down creating the fold shape (see Sec.~\ref{sec:iso-value-estimation}).

Solving problem (1) is much simpler and different interpolation methods and surface representations could be used.
In this work we opted for using Radial Basis Functions (RBF, see Sec.~\ref{sec:rbf}) interpolation since it provides good results and uses the same mathematical tools required by the HBRBF interpolation.
In fact, RBF interpolation can be seen as a special case of HBRBF interpolation where the only constraint type are points with values.
However, we do not define the terrain as an implicit surface of a potential-field as was the case of the rock layers.
Instead, the terrain is modeled as a height-map, where we know the height of the 2D points sampled along the contour curves of the topographic map.
Therefore, these points are mapped as points of the RBF function, and the height as the value.
In this way, we can explicitly evaluate the height of any point on the 2D plane.
This is useful because it provides an efficient way to find the height of points from contacts and symbols defined on the geological map sketch.

To solve problem (3) and to create the final geological model, we combine the implicit surfaces using Constructive Solid Geometry (CSG, see Sec.~\ref{sec:csg}).
We also use CSG to create different visualizations cuts defined by user sketches.
Then, to visualize the created model we triangulate it using Marching Tetrahedra~\cite{treece1999}.
To fit the terrain's height-map representation in this CSG framework we first need to transform it into an implicit surface so we can easily decide whether a point is inside of the terrain.
This is done by simply defining it as the iso-surface of iso-value equal to zero, of a function of the distance to the terrain, where points below the terrain have negative distance.
Finally, we define the boundaries using the HBRBF implicit formulation to create five extra planes to close the sides and bottom of the 3D geological model.

\begin{figure}[hb!]
  \centering
  \includegraphics[width=0.95\linewidth]{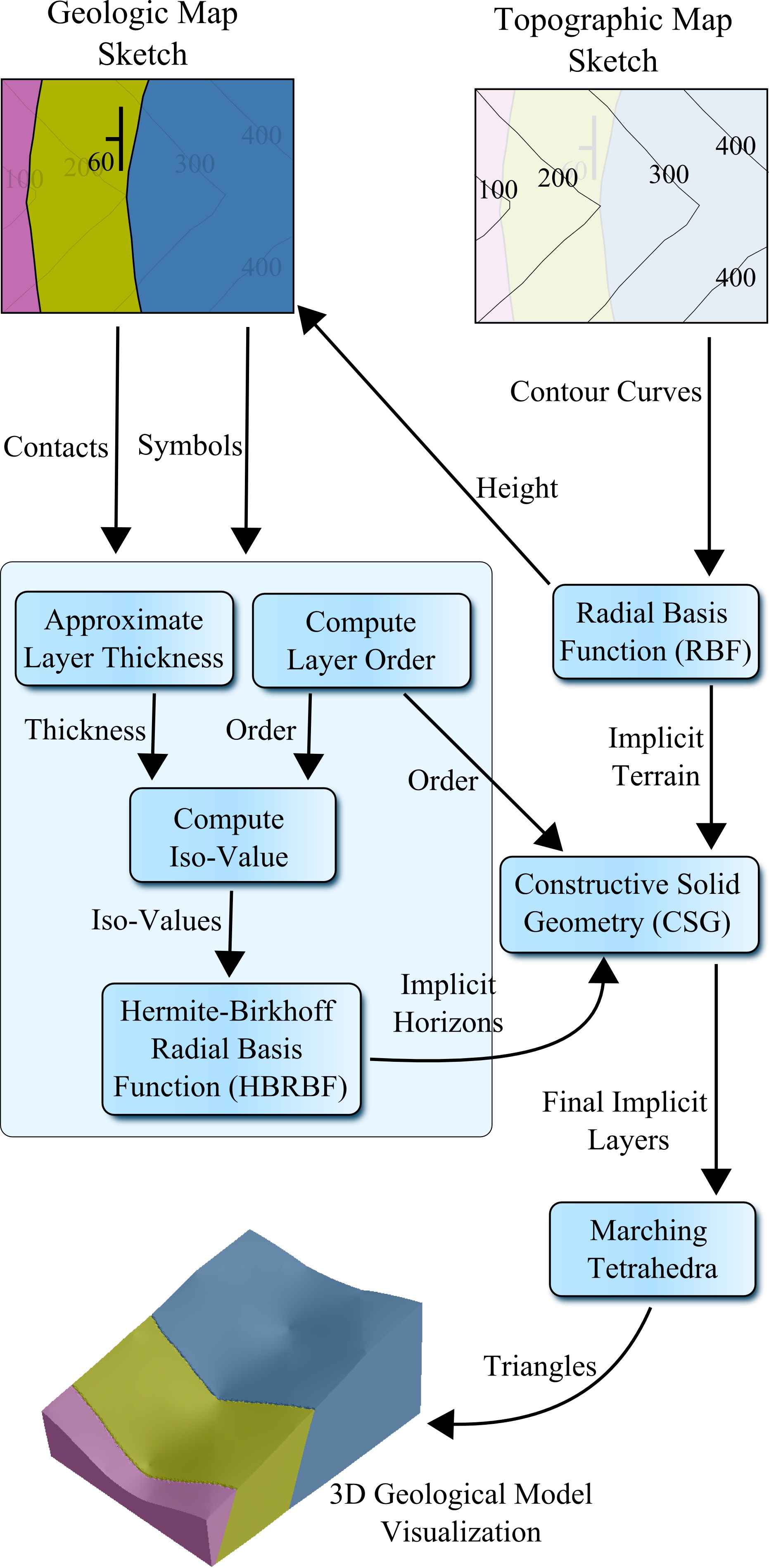}
  \caption{Workflow for creating a 3D geological model from topographic and geological map sketches.
  The topographic map defines contour curves that are interpolated using RBF.
  The RBF terrain is used to find the height of symbols and contacts on the geological map.
  The contacts and symbols on the geological map are used to approximate the thickness of the layers, find their order, and create implicit definition of the rock series and layers interfaces using HBRBF.
  The CSG then combines the order information, and the horizon and terrain implicits to define implicitly each rock layer.
  Finally, all layers are visualized by the Marching Tetrahedra triangulation.} 
  \label{fig:workflow}  
\end{figure}

Fig.~\ref{fig:workflow} illustrates the entire workflow for the 3D geological model creation from topographic and geological map sketches.
The next sections present the mathematical details of how topographic map sketches are converted into a 3D representation using RBFs, how HBRBFs are used to interpolate the geological map, and finally how CSG is used to create the final 3D model.

\subsection{Radial Basis Functions (RBFs)}
\label{sec:rbf}

In this section we describe the method used to create the 3D terrain surface from the set of sparse samples provided by the topographic map sketch.
As previously described, a sketched topographic map defines a set of curves that are discretized into points $\vec p^i \in \R^3$, where $p^i_1$ and $p^i_2$ correspond to coordinates on a plane representing the topographic map, and $p^i_3$ the only height value of the terrain at $(p^i_1, p^i_2)$.
Based on these assumptions we can model the terrain's height $p_3$ at any point on the topographic map, as a function $f: \R^2 \rightarrow \R$ such that $p_3 = f(p_1, p_2)$.
All we need now is to find a function $f^*: \R^2 \rightarrow \R$ that approximates $f$ by interpolating all samples $p^i$ provided.
In a more formal description, let $\overline S = \{\vec p^1, \vec p^2, \vec p^3, \cdots, \vec p^N\}$ be the set of all samples provided by the discretization of the contour curves of the topographic map. 
We need to find a function $f^*(p_1, p_2) = p_3\;\; \forall\; \vec p \in \overline S$.

In this work we use Radial Basis Functions (RBF) interpolation to construct $f^*$.
The RBF interpolation is a multi-variable scheme for function interpolation and is defined as:

\begin{equation}
f^{*}(\vec{x}) = \displaystyle\sum_{i=1}^{N} \alpha_i \phi(\norm {\vec{x}-\vec{p}^i}) 
\end{equation}
where $\alpha_i$ are real coefficients that we would like to find, $N$ the number of samples $\vec p^i$ where the value of $f(\vec p^i)$ is known. $\phi: \R_+ \rightarrow \R$ are given continuous functions called radial basis function kernel, and the norm argument $\norm {\cdot}$ is usually taken as the Euclidean norm.
To find the values of $\alpha_i$ and construct the interpolant $f^*(\vec x)$, it is necessary to solve a linear system $A\vec \alpha = \vec b$.
Each row of the matrix and entry in the $\vec b$ vector corresponds to a know value of $f$.

\begin{equation}
 \begin{pmatrix}
  \phi_{1,1} & \phi_{1,2} & \cdots & \phi_{1,N} \\
  \phi_{2,1} & \phi_{2,2} & \cdots & \phi_{2,N} \\
  \vdots  & \vdots  & \ddots & \vdots  \\
  \phi_{N,1} & \phi_{N,2} & \cdots & \phi_{N,N} \\ 
 \end{pmatrix} 
 \begin{pmatrix}
  \alpha_1\\
  \alpha_2\\
  \vdots \\
  \alpha_N
 \end{pmatrix}
 =
 \begin{pmatrix}
  v_1\\
  v_2\\
  \vdots \\
  v_N
 \end{pmatrix},
\end{equation}
where $\phi_{i,j} = \phi(\norm {\vec p^j - \vec p^i})$, and $v_i$ the value of the function at $\vec p^i$.
The RBF kernel $\phi$ can have different definitions depending on the application.
However, to guarantee the unique solvability of the system presented, RBF kernels must satisfy some conditions. In this article we will not discuss these conditions, however, the reader interested can find the details in~\cite{buhmann2003}.
For interpolating the terrain defined by the topographic map we use the Norm kernel $\phi(r) = r$.

\begin{figure}[ht!]
    \centering
    \includegraphics[width=\linewidth]{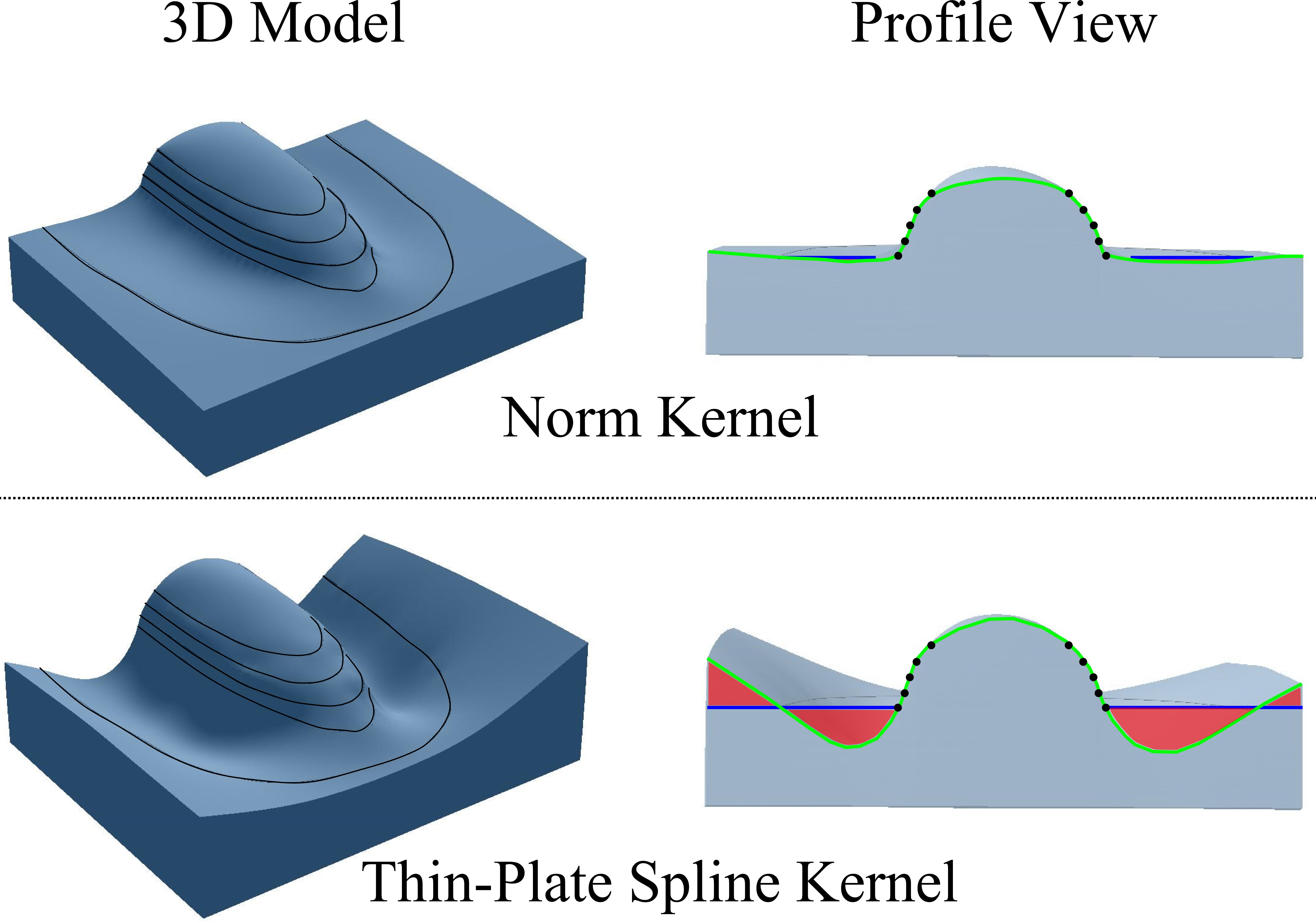}
    \caption{Comparison between two different kernel types: norm, and thin-plate spline. The norm kernel on top yelds the best results. On the right-side the models' back profile show the actual surface in green and the problematic area in red for each kernel type.}
    \label{fig:rbf_kernels}  
  \end{figure}

A polynomial term $P \in \mathds{P}^{1}$ can be introduced in the RBF interpolation definition as described below:
\begin{empheq}{alignat=1}
 f^{*}(\vec{x}) =& \displaystyle\sum_{i=1}^{N} \alpha_i \phi(\norm {\vec{x}-\vec{p}^i}) + P(\vec x) \label{eq:rbf}\\
 0 = & \displaystyle\sum_{i=1}^{N} \alpha_i  \vec p^i,\label{eq:rbf-sc-poly-const}\\
 0 = & \displaystyle\sum_{i=1}^{N} \alpha_i \label{eq:rbf-sc-poly-var},
\end{empheq}
where Eq.~\ref{eq:rbf} defines the new interpolant $f^*$ with the polynomial term $P$.
And Eq.~\ref{eq:rbf-sc-poly-const}~and~\ref{eq:rbf-sc-poly-var} define the side conditions necessary to take the extra degrees of freedom introduced by the polynomial term.
In our work we choose a polynomial term of first degree, such that $P(\vec x) = b + <\vec a, \vec x>$ which allows our interpolant $f^*$ to represent planes exactly.
The polynomial term also relax some conditions of the RBF kernels, enabling the use of different kernels.
With the introduction of the first order polynomial term the new linear system can be expressed in the matrix form as:
\begin{equation}
\resizebox{\hsize}{!}{$
\setlength{\arraycolsep}{3pt}
 \begin{pmatrix}
  \phi_{1,1}       & \phi_{1,2}         & \cdots & \phi_{1,N} &  \vec P(\vec p^1)^T\\
  \phi_{2,1}       & \phi_{2,2}         & \cdots & \phi_{2,N} &  \vec P(\vec p^2)^T\\
  \vdots           & \vdots             & \ddots & \vdots  \\
  \phi_{N,1}       & \phi_{N,2}         & \cdots & \phi_{N,N}         &  \vec P(\vec p^N)^T\\ 
  \vec P(\vec p^1) & \vec P(\vec p^2)   & \cdots & \vec P(\vec p^N)   & \matrixinside{0}_{(d+1)\times (d+1)} \\  
 \end{pmatrix} 
 \begin{pmatrix}
  \alpha_1\\
  \alpha_2\\
  \vdots \\
  \alpha_N\\
  b\\
  \vec a
 \end{pmatrix}
 =
 \begin{pmatrix}
  v_1\\
  v_2\\
  \vdots \\
  v_N\\
  0\\
  \vec 0\\
 \end{pmatrix},
 $}
\end{equation}
where $\vec P(\vec x^i) = [1\;x_1^i\;x_2^i\;\cdots\;x_d^i]^T$, being $d$ the dimension of the point, which is $2$ for our topographic map interpolation.
By solving this system we are able to reconstruct the terrain by interpolating the contour curves.
So, to find the height of the terrain at any $(x_1,x_2)$ we just need to evaluate $f^*(x_1,x_2)$ using the parameters found by solving the linear system.
As previously stated, in this work, we choose the Norm kernel, which is a particular case of the multiquadrics RBF kernel ($\phi_{MQ}(r) = \sqrt{c + r^2}$), with $c=0$, such that $\phi(r) = \sqrt{0 + r^2} = r$.
Different kernels yields different results, which are more or less suitable to reproducing the 3D terrain represented by the topographic map.
Fig.~\ref{fig:rbf_kernels} illustrates 2 different choices of kernel: norm, and thin-plate spline.
The thin-plate spline kernel is defined as $\phi_{TP}(r) = r^2 \ln(r)$.
It is easy to observe that the thin-plate spline kernel is not the best choice.
The 3D terrain constructed has heights that are lower than the value between the two contour curves it is within, therefore not represented in the topographic map sketched.
The norm kernel though, have the tendency be flatter between two consecutive contour curves and result in a better approximation of the topographic map.

The described approach explicitly gives the height value for any $(x_1,x_2)$, which is desirable for efficiently finding the height of the constraints of the geological map.
However, to fit in our implicit surface framework for constructing the 3D geological model using CSG, we need to transform our function $f^*$ into a function $F^*: \R^3 \rightarrow \R$ such that the terrain is defined as the implicit surface $S_f = \{ \vec x \in \R^3 \mid F^*(\vec x) = 0\}$.
Which is accomplished by making $F^*(x_1, x_2, x_3) = x_3 - f^*(x_1, x_2)$.
So that $F^*(x_1,x_2,x_3) \leq 0$ for points inside the terrain surface, i.e., in the subsurface, and $F^*(x_1,x_2,x_3) > 0 $ otherwise.

\subsection{Hermite-Birkhoff Radial Basis Functions (HBRBF)}
\label{sec:hbrbf}



Differently from the terrain interpolation approach using RBF, we interpolate the rock layers interfaces based on the contacts-graph and symbols directly as an implicit surface $S_g$ of $g:\R^3 \rightarrow \R$ where $S_g = \{ \vec x \in \R^3 \mid g(\vec x) = l_h\}$.
Geological maps contacts and symbols provide 3 different types of interpolation constraints.
(1) Contacts are discretized in a set of points $\vec p^i \in \R^3$ such that each point $\vec p^i$ in the contacts-graph provides a constraint $g(\vec p^i) = l_h$, where $l_h$ is the iso-value associated with a horizon $h$ it is interpolating (details on how to select iso-values for the layers are discussed in Section~\ref{sec:iso-value-estimation}).
(2) Dip-and-strike symbols are a set of points $\vec q^j \in \R^3$ with information about the dipping of the rock layer.
This dipping information provides constraints $\nabla g(\vec q^j) = \vec n^j_q$ where $\vec n^j_q$ are gradients orthogonal to the dipping direction of the symbol at $\vec q^j$.
(3) Anticlines, synclines, domes and basins symbols are a set of points $\vec u^k \in \R^3$, at the symbols center or strike-curve, with information about the folding of the rock series.
This information provides two constraints for the same point $\vec u^k$ such that $g(\vec u^k) = z_k$ and $\nabla g(\vec u^k) = \vec n^k_u$, where $n^k_u$ are gradients $(0, 0, 1)$, and  $z_k$ are specifically chosen to deform $g(x)$ and create the up or down folds.

We model each rock series as a single function $g$, therefore, rock layers belonging to a same rock series are sub-parallel to each other and do not contain intersections, which enforces part of geological rules previously stated.
Therefore, we want to find a function $g^*:\R^3 \rightarrow \R$ that approximates $g$ such that:
\begin{flalign} 
 g^*(\vec p^i) &= l_h &\;\;\;\text{\tiny points with values} \\
\nabla g^*(\vec q^j) &= \vec n_q^j &\;\;\;\text{\tiny points with gradients}\\
g^*(\vec u^k) = z_k,\;\; g^*(\vec u^k) &= \vec n_u^k&\;\;\;\text{\tiny points with values and gradients}
\end{flalign}
The Hermite-Birkhoff Radial Basis Function (HBRBF) interpolation introduced by Macedo~\cite{macedo2011} meet such requirements and is given by:
\begin{flalign} 
  g^{*}(&\vec{x}) = \underbrace{\displaystyle\sum_{i=1}^{N_v} \alpha_i \phi(\norm {\vec{x}-\vec{p}^i})}_\text{Value Interp.} \underbrace{- \displaystyle\sum_{j=1}^{N_g} \langle \vec \beta^j, \nabla \phi(\norm {\vec{x}-\vec{q}^j}) \rangle}_\text{Gradient Interp.} \\
         \nonumber& \underbrace{+  \displaystyle\sum_{k=1}^{N_{vg}}\{ \gamma_k \phi(\norm {\vec{x}-\vec{u}^k}) - \langle \vec \lambda^k, \nabla \phi(\norm {\vec{x}-\vec{u}^k}) \rangle \}}_\text{Value \& Gradient Interp.} + \underbrace{P(\vec x)}_\text{Poly.}
\end{flalign}
where $N_v$, $N_g$, $N_{vg}$ are the number of points with values, gradients, and values \& gradients, respectively. 
$\alpha$, $\vec \beta$, $\gamma$, and $\vec \lambda$ are the unknowns.
$\nabla \phi(r)$ is the gradient of the RBF kernel $\phi(r)$.
The $\nabla g^*(x)$ which is necessary to interpolate the gradient information for points $\vec q^j$ and $\vec u^k$ is given by:
\begin{flalign}
   &\nabla g^*(\vec x) = \displaystyle\sum_{i=1}^{N_v}     \alpha_i \nabla \phi(\norm {\vec{x}-\vec{p}^i}) - \displaystyle\sum_{j=1}^{N_g}  H \phi(\norm {\vec x-\vec q^j}) \vec \beta^j  \\
          \nonumber& +    \displaystyle\sum_{k=1}^{N_{vg}} \{ \gamma_k \nabla \phi(\norm {\vec{x}-\vec{u}^k}) -                             H \phi(\norm {\vec x-\vec u^k}) \vec \lambda^k \} + \nabla P(\vec x),
\end{flalign}
where $H$ is the Hessian, a $3\times 3$ matrix of second-order partial derivatives of $\phi(\vec x)$.

For being able to uniquely solve the system arising from these equations we also need the following side conditions introduced by the use of the polynomial term:

\begin{empheq}[left=\empheqlbrace]{alignat=3}
 0 = & \displaystyle\sum_{i=1}^{N_v} \alpha_i + \displaystyle\sum_{k=1}^{N_{vg}} \gamma_k \label{eq:sc-poly-constant}\\
 0 = & \displaystyle\sum_{i=1}^{N_v} \alpha_i  \vec x^j +\displaystyle\sum_{j=1}^{N_{g}} \vec \beta_j + \displaystyle\sum_{k=1}^{N_{vg}} (\gamma_k \vec x^k + \vec \lambda_k)
\end{empheq}
This formulation yields the following linear system to be solved:
\begin{equation}
 \begin{pmatrix}
   \matrixinside{A}_{N_v \times T} \\
   \matrixinside{B}_{3N_g \times T} \\
   \matrixinside{C}_{4N_vg \times T} \\
   \matrixinside{D}_{4 \times T}
 \end{pmatrix}
    \begin{pmatrix}
      \matrixinside{\alpha}\\
      \matrixinside{\beta}\\
      \matrixinside{\gamma}\\
      \matrixinside{\lambda}\\
      b\\
      \vec a
    \end{pmatrix}
    =
    \begin{pmatrix}
    \vec v_v\\
    \vec v_g\\
    \vec v_{vg}\\    
    0 \\
    \vec 0 \\
  \end{pmatrix},
\end{equation}
where $T = N_v + 3N_g + 4N_{vg} + 4$, and:
\begin{equation}
\resizebox{\hsize}{!}{$
\vec v_v=
\begin{pmatrix}
g^*(\vec p^1) \\
g^*(\vec p^2) \\ 
\vdots \\
g^*(\vec p^{N_v})
\end{pmatrix},
\vec v_g =
\begin{pmatrix}
\nabla g^*(\vec q^1) \\
\nabla g^*(\vec q^2) \\ 
\vdots \\
\nabla g^*(\vec q^{N_g})
\end{pmatrix},
\vec v_{vg} =
\begin{pmatrix}
g^*(\vec u^1) \\
\nabla g^*(\vec u^1) \\ 
\vdots \\
g^*(\vec u^{N_g})\\
\nabla g^*(\vec u^{N_g})
\end{pmatrix},
$}
\end{equation}
such that $\vec v_v$, $\vec v_g$, and $\vec v_{vg}$, are the known values, gradients, and values \& gradients, respectively at points $\vec p^i$, $\vec q^j$, and $\vec u^k$.
The block matrices $A$, $B$, $C$ are related to these same known values, gradients and values \& gradients.
Such that each sub-block $A_i$, $B_j$, and $C_k$ represents a single constraint, and $D$ is the the block matrix representing the side conditions, such that:
\begin{flalign}
\resizebox{0.85\hsize}{!}{$
\matrixinside{A}_i =
\begin{pmatrix}
 \matrixinside{\phi}^{\overline v}(\vec p^i)           & -\matrixinside{\nabla \phi}^{\overline g}(\vec p^i)  &  \matrixinside{\phi}^{\overline {vg}}(\vec p^i) & -\matrixinside{\nabla \phi}^{\overline {vg}}(\vec p^i)  & \vec P(\vec p^i)^T \\
 \end{pmatrix}
 $} \\
 \resizebox{0.85\hsize}{!}{$
\matrixinside{B}_j =
\begin{pmatrix}
  -\matrixinside{\nabla \phi}^{\overline v}(\vec q^j)  & -\matrixinside{H}^{\overline g}(\vec q^j)            &  \matrixinside{\nabla \phi}^{\overline {vg}}(\vec q^j)       & -\matrixinside{H}^{\overline {vg}}(\vec q^j)  & \vec P(\vec q^j)^T \\ 
\end{pmatrix}
$}\\
\resizebox{0.85\hsize}{!}{$
\matrixinside{C}_k =
\begin{pmatrix}
    \matrixinside{\phi}^{\overline v}(\vec u^k)             &  \matrixinside{\nabla \phi}^{\overline g}(\vec u^k)  &  \matrixinside{\phi}^{\overline {vg}}(\vec u^k)         & -\matrixinside{\nabla \phi}^{\overline {vg}}(\vec u^k) & \vec P(\vec u^k)^T \\
   -\matrixinside{\nabla \phi}^{\overline v}(\vec u^k) & -\matrixinside{H}^{\overline g}(\vec u^k)            & -\matrixinside{\nabla \phi}^{\overline {vg}}(\vec u^k)       & -\matrixinside{H}^{\overline {vg}}(\vec u^k)           & (\matrixinside{\nabla P}^{k})^T 
\end{pmatrix}
$}\\
\resizebox{0.85\hsize}{!}{$
\matrixinside{D} =
\begin{pmatrix}
 \vec P(\vec p^{\overline v}) & \vec \nabla P(\vec q^{\overline g}) & \vec P(\vec u^{\overline {vg}}) & \vec \nabla P(\vec u^{\overline {vg}}) & \vec 0_{4 \times 4} \\
\end{pmatrix}
$}
\end{flalign}
where superscripts $\bar v$, $\bar g$, and $\bar {vg}$, in the block matrices, indicate the sets of all points $\vec p^i$, $\vec q^j$, and $\vec u^k$, respectively.
The vector $\matrixinside{\phi}^{\vec \cdot}(\vec x) = \phi(\norm {\vec{x}-\vec \cdot})$, and the matrices $\matrixinside{\nabla \phi}^{\vec \cdot}(\vec x) = \nabla \phi(\norm {\vec{x}-\vec \cdot})$, and $\matrixinside{H}^{\vec \cdot}(\vec x) = H \phi(\norm {\vec{x}-\vec \cdot})$
Such that $\matrixinside A_i$ is a $1 \times T$ matrix, $\matrixinside B_j$ a $3 \times T$ matrix, $\matrixinside C_k$ a $4 \times T$ matrix, and $D$ a $4 \times T$ matrix.

The linear system described, which is composed of a symmetric indefinite matrix, can be solved using a $LDL^T$ decomposition.
Finally, for this application we use the Duchon kernel $\phi(r) = r^3$, which yields the best results for our application.
Moreover, the number of choices is restricted since the kernel has to be $C^2$ continuous because of the Hessian.

\subsection {Iso-value Estimation}
\label{sec:iso-value-estimation}

\begin{figure}[ht!]
  \centering
  \includegraphics[width=\linewidth]{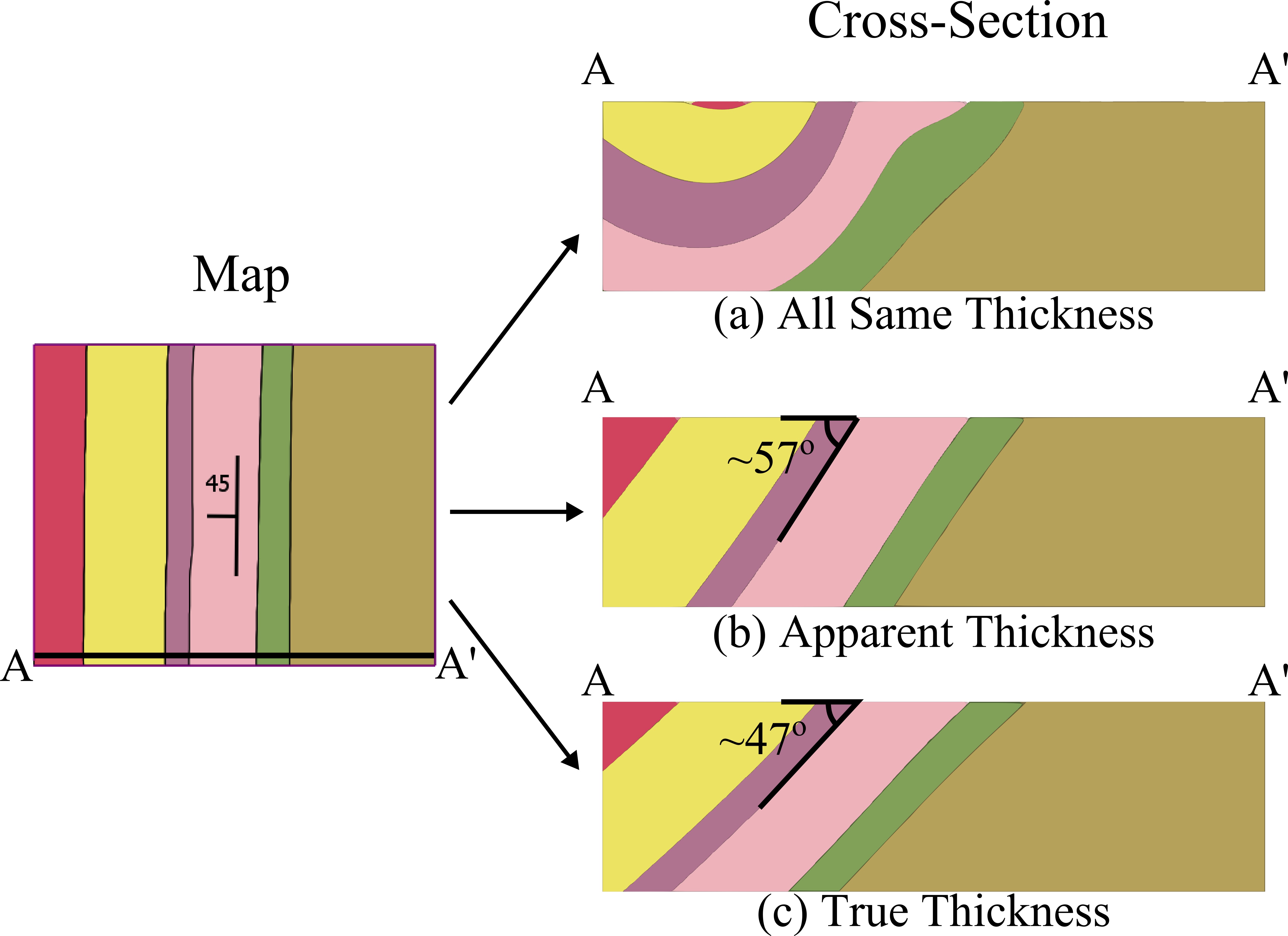}
  \caption{Comparison between approaches to compute rock layer thickness.
  (a) Assuming all layers have the same thickness. The resulting model is not what is expected from the map.
  (b) Using apparent thickness, the results make sense visually but the dipping angle is far from the defined on the map.
  (c) Using true thickness estimation the results and angle is much closer to what is represented on the map.}
  \label{fig:thickness}  
\end{figure}

To select the iso-values for the rock layers interfaces the first step is to note that rock layers are stacked on top of each other with older rocks on the bottom.
Based on this we can select the iso-values such that older rock layer interfaces have iso-values lower than younger interfaces.
There are three possibilities to estimate the iso-value for each layer interface.
The first is to assume all layers have the same thickness and distribute iso-values uniformly spaced on the layers.
However, this approach only works when all layers have approximately the same thickness.
Another possibility is to use the apparent thickness of the layer, which is one calculated based only on the thicknesses of the layers on the top of the 2D geological map.
This approach solves part of the problem of layers of different thicknesses.
However it also has its problems, for instance, imagine a rock layer dipping in a very low angle close to horizontal, the thickness on top of the map is much greater than the true thickness.
Fig.~\ref{fig:thickness}(a) and (b) illustrates both approaches.
To appropriately interpolate the rock layers using the HBRBF method, it is important to better estimate the thickness of each rock layer.
The true thickness is then calculated by combining the apparent thickness with the dipping of the layers.
Fig.~\ref{fig:thickness} presents a comparison between all approaches showing how the use of the true thickness improves the results.

To find the best rock thickness for our sketch we would need the 3D model, which is what where are actually trying to construct.
Therefore, we need to estimate the true thickness based only on the existing geological maps sketch.
To this end, we use the apparent thickness and the dipping symbols to try to estimate the true thickness of the layer.
We basically, make an average of the existing dipping angles in the series the rock layer belongs.

To calculate the apparent thickness efficiently we first compute the area occupied by the layer on the geological map and divide it by the length of the horizon also on the geological map.
Fig.~\ref{fig:thickness}(b) presents the result of using the apparent thickness for the layer.
The layer dipping angle in the figure is approximately $57^o$ which is considerably different from the $45^o$ expected.
The true thickness is then computed using the apparent thickness and the average dipping angle provided by the symbols.
The true thickness is given by the formula $T_t = T_a sin(\theta)$, where $T_t$ is the true thickness, $T_a$ the apparent thickness and $\theta$ the average dipping angle.
Fig.~\ref{fig:thickness}(c) presents the result when using the true thickness estimation.
The resulting angle reproduced is about $47^o$, which is relatively close to the $45^o$ specified.
However, we need to also account for the terrain effect.
When the terrain slope is high the apparent thickness is much smaller than the thickness shown by the vertical axis, which is called vertical thickness.
In such cases the vertical thickness is used instead of the apparent thickness, and the formula to compute the true thickness is $T_t = T_v cos(\theta)$, where $T_v$ is the vertical thickness.

Finally, we can use the iso-values of the rock layers interfaces to create folds such as anticlines, synclines, domes and basins.
As previously described in Sec.~\ref{sec:creating-3d-geological-model}, these types of folds symbols interpolate values \& gradients at points in their center or strike-curve.
The gradient we use to construct these folds is always a vector pointing up $(0,0,1)$.
For creating up-folds such as anticlines and domes, the values at these points are chosen to be lesser than the iso-value of the rock layer interface (contact) which is enclosing the symbol on the geological map, and above the immediately older interface.
For down-folds such as synclines and basins, these values are chosen to be greater than the iso-value of the enclosing contact, and lesser than the immediately younger interface.

\subsection{Faults}
\label{sec:faults-construction}
\begin{figure}[ht!]
  \centering
  \includegraphics[width=0.8\linewidth]{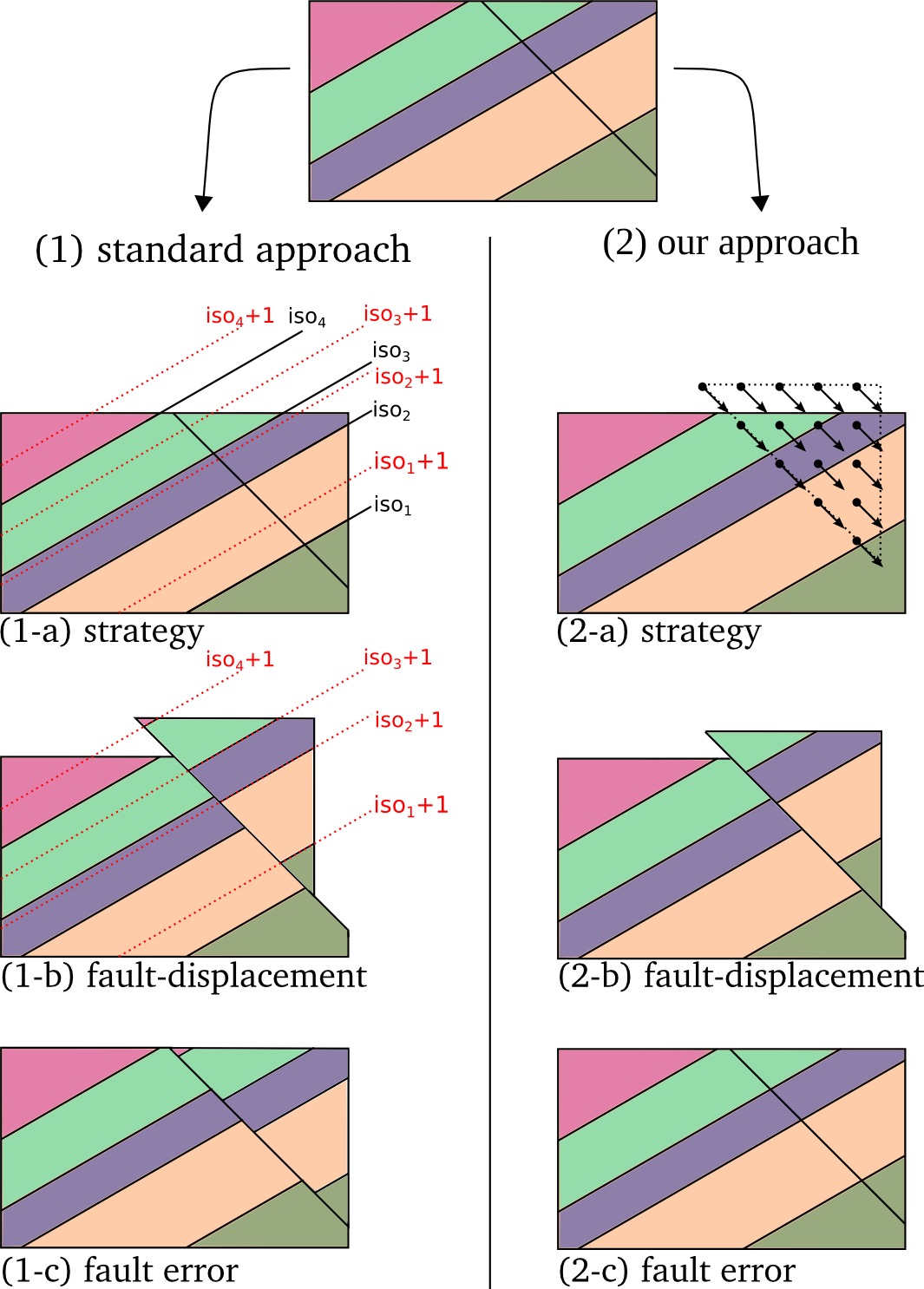}
  \caption{Comparison between standard approach for creating faults using implicits.
  (1) the standard approach creates artificial results and introduces errors.
  (2) our approach is more computationally expensive but the problems with the standard approach are avoided.}
  \label{fig:fault-displacement}  
\end{figure}

Faults are surfaces representing discontinuities in geological layers caused by external forces that break and displace the rock.
In order to model this discontinuity the standard method is to simply add a constant to the interpolant $g^*(\vec x)$ in one side of the implicit fault surface $F$~\cite{marechal1984}, such that:
\[ G^*(\vec x)  = \begin{cases} 
      g^*(\vec x) + c & \textrm{ if $\vec x \in F^+$} \\
      g^*(\vec x)     & \textrm{ if $\vec x \in F^-$},
   \end{cases} \]   
where $c \in \R$, is the constant defining the fault displacement.
$F^+$ and $F^-$ two different sides of the fault surface $F$.
The problem with such approach is that it actually transforms a level-set into another, so that the shape is not preserved.
Moreover, the displacement also does not follow the fault dipping direction but the interpolant's gradient.
Fig.~\ref{fig:fault-displacement}(1) illustrates an example of the error that occurs when using this method.

\begin{figure}[ht!]
\centering
  \includegraphics[width=\linewidth]{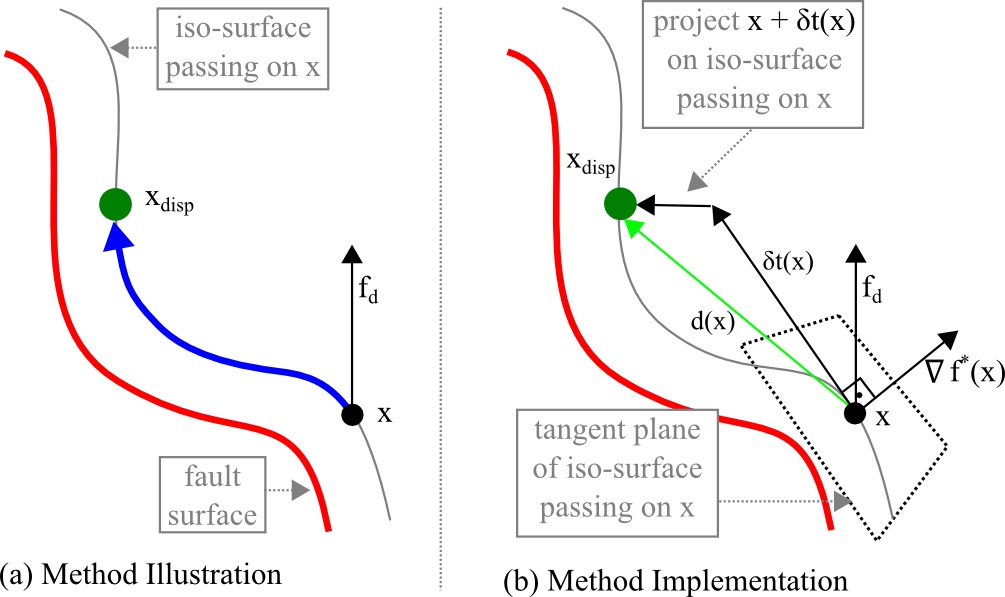}
  \caption{Cross-section illustration of the modeling of faults.
  In (a) a high-level description of the displacement of points in one side of the fault.
  In (b) the method implementation as described in Eq.~\ref{eq:fault-modeling-1}~through \ref{eq:fault-modeling-4}.}  
  \label{fig:fault-methodology}  
\end{figure}

To solve the aforementioned problems, we propose a different approach (see Fig.~\ref{fig:fault-displacement}(2)).
The general idea is to create the discontinuity by ``sliding'' points on one side of the fault surface following the fault surface shape on the opposite direction of the fault displacement (Fig.~\ref{fig:fault-methodology}(a)).
So, instead of adding a constant to $g^*$ in one side of the fault surface, we evaluate the composition $g^*(M(\vec x, \vec f_d, \delta))$, where $M: \R^7 \rightarrow \R^3$ is a mapping given by:
\begin{equation}
M(\vec x, \vec f_d, \delta)  = \begin{cases} 
      \vec x + d(\vec x, \vec f_d, \delta) & \textrm{ if $\vec x \in F^+$} \\
      \vec x             & \textrm{ if $\vec x \in F^-$},
   \end{cases}
   \label{eq:fault-modeling-1}
\end{equation}
and $d: \R^7 \rightarrow \R^3$ yields the displacement vector at point $\vec x$, $\vec f_d$ and $\delta$ are the given fault displacement direction and amount specified by the fault symbol sketched, respectively.
The computation of $d(\vec x, \vec f_d, \delta)$ is detailed next and illustrated in Fig.~\ref{fig:fault-methodology}(b).
To compute the displacement vector, we first displace $\vec x$, in the opposite direction of $f_d$, on the tangent plane of the iso-surface where $\vec x$ is.
Then, we project it back to the same iso-surface (see Fig.~\ref{fig:fault-methodology}).
In more details, let $f^*: \R^3 \rightarrow \R$ be the function interpolating the implicit fault surface $F = \{\vec x \in \R^3 \mid f^*(\vec x) = 0\}$.
Where $f^*$ is an HBRBF that interpolates the fault symbol dipping and trace on the geological map (HBRBF interpolating points with values \& gradients).
To compute the displacement vector $d(\vec x, \vec f_d, \delta)$ for a point $\vec u \in F^+$ the idea is to project the opposite of the fault direction $-\vec f_d$ (up $(0, 0, 1)$ or down $(0, 0, -1)$, for reverse or normal faults) on the tangent plane of the iso-surface of $f^*(\vec x)$ at $\vec u$.
To do so, we first compute $\nabla f^*(\vec u)$.
Then, based on the opposite fault direction vector $-\vec f_d$, we compute a rotation axis $\vec r = \nabla f^*(\vec u) \times \vec f_d$ and rotate $\nabla f^*(\vec u)$ around $\vec r$ by $90^\circ$. 
The unit vector of the rotated vector, which we call $t(\vec u, \vec f_d)$, gives the direction of the displacement on the tangent plane, this unit vector is finally multiplied by the displacement amount $\delta$.
The last step is to project $\vec u + \delta t(\vec u, \vec f_d)$ back on the iso-surface of $f^*$ which $\vec u$ was initially on.
In this work we use a gradient descent method to find $\vec u + d(\vec u, \vec f_d, \delta)$ from $u + \delta t(\vec u, f_d)$, with a maximum of $10$ iterations or error $e_{abs} = (f^*(\vec u ) - f^*(\vec u_{new}))^2 \leq 10^{-8}$. 
In summary:
\begin{align}
d(\vec x, \vec f_d, \delta) & = 
\begin{cases}
P_{proj}(\vec x + \delta t(\vec x, \vec f_d)) & \textrm{ if $r(\vec x) \neq \vec 0$} \\
\vec 0 & \textrm{otherwise}\\
\end{cases} \\
t(\vec x, \vec f_d) &= \frac{R(r(\vec x)) \vec f_d}{R(r(\vec x)) \vec f_d} \\
r(\vec x) &= \nabla f^*(\vec x) \times \vec f_d, \label{eq:fault-modeling-4}
\end{align}
where $R(\vec r)$ is the rotation matrix around axis $\vec r$, and $P_{proj}(\vec p)$ projects a point $\vec p$ on the iso-surface of $f^*$ that passes on $\vec x$. 


If more accuracy in the fault displacement is required, the method described can be applied $n$ times with fractions $\delta/n$, and approximate better the displacement of points on the geodesic of the iso-surface.
However, there is a trade-off between the accuracy and its computational cost.
The results presented in this article used only a single step, i.e., $n=1$.
Finally, another advantage of the proposed fault modeling approach is that we can specify the direction of faults by any vector, which allows the modeling of strike-slip faults, and even oblique-faults, which are a combination of strike-slip and dip-slip faults.

\subsection{Constructive Solid Geometry (CSG)}
\label{sec:csg}

Geologic models represent the 3D disposition and geometry of rocks in the subsurface.
Therefore, each rock layer is a solid (volume) and where one rock layer ends another one starts without gaps or holes.
The HBRBF and RBF formulations alone, presented in the previous sections, enable the representation of the geological horizons surfaces only.
However, we are interested in constructing a water-tight 3D model where each rock layer is a separate volume.
In order to achieve this goal we take advantage of the implicit representation of the surfaces and use Constructive Solid Geometry (CSG) to build all rock series and their respective rock layers by combining the implicit representations interpolating the rock series and layers interfaces.
In the implicit representation used we define a point $P$ to be inside the solid when $f^*(P) < 0$.
Using this concept of inside and outside, CSG enables us to create the actual rock layers by combining the implicit representations using set operators, such as intersection and subtraction.

\begin{figure}[ht!]
  \centering
  \includegraphics[width=\linewidth]{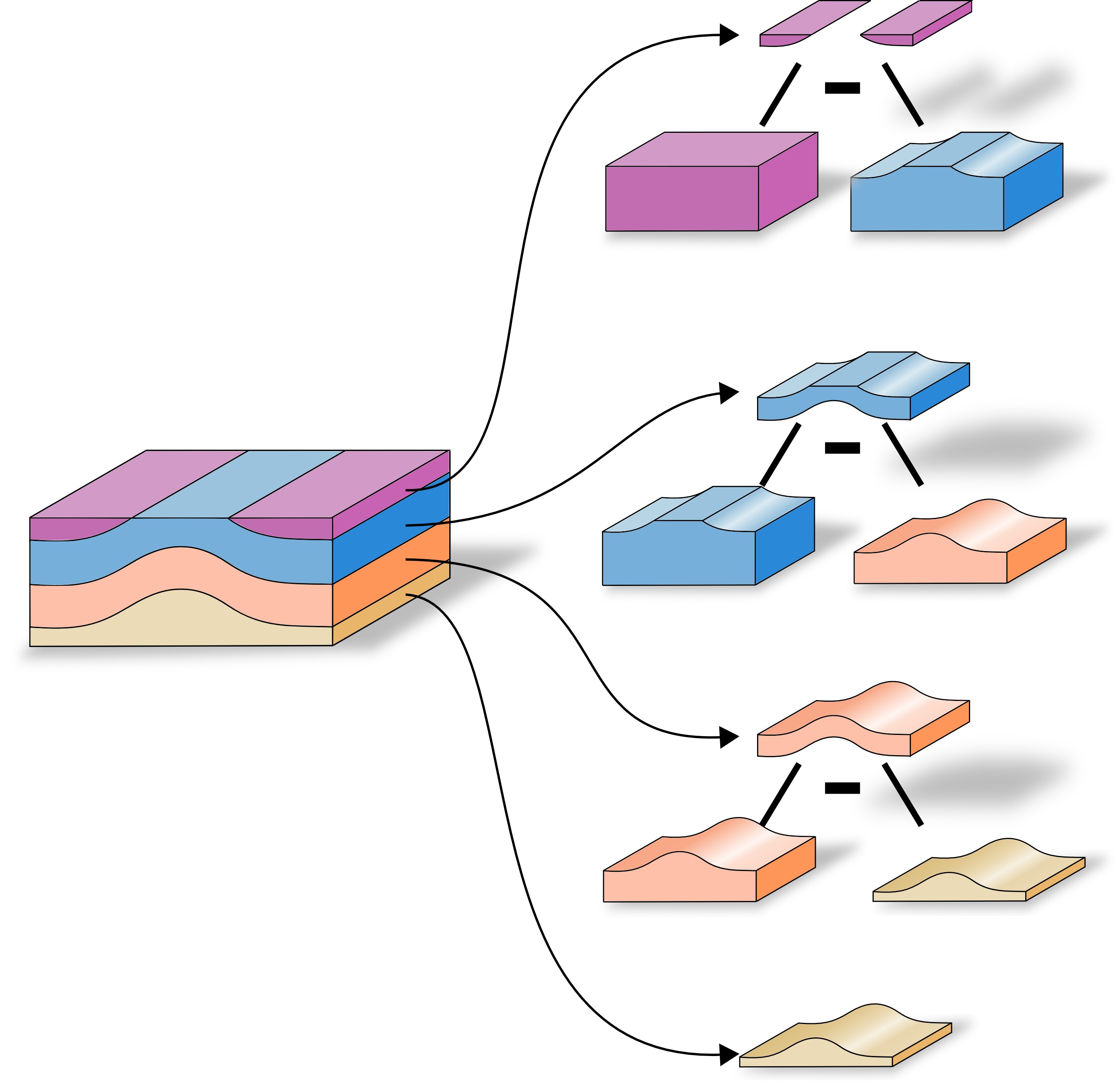}
  \caption{Illustration of the CSG for creating the water-tight rock layers.
  Each horizon implicit is subtracted by the immediately older horizon.}
  \label{fig:csg}  
\end{figure}

To create a 3D solid representing the rock layer we simply subtract the implicit representation of the end of this rock layer with the end of the previous rock layer.
In Fig.~\ref{fig:csg} we illustrate this process by showing the construction of different rock layers of the same rock series using this idea.
For incorporating the effects of another rock series, the older series is intersected with the surface of unconformity, and is subtracted by the newer series on its top.
Using this simple approach all rock layers can be constructed as solids creating valid geological models where each series and layer fits perfectly on top of the other.

This same CSG approach can be used to enable the inspection of the 3D model interior through cross-sections defined by sketches on the 2D map.
The idea is to create a cutting surface out of the sketch such that everything on the left side of the sketch is removed from the 3D model.
This cutting surface is created using the HBRBF formulation with all points in the sketch being points with values \& gradients.
Where the gradients are vectors orthogonal to the sketch curve on the 2D map plane pointing to the left side of the curve.
Finally, this surface is combined with the 3D geological model using the CSG ``and'' set operation creating the cut solid model.

\section{Results and Discussions}
\label{sec:results_and_discussions}

\begin{figure*}[ht!]
  \centering
  \includegraphics[width=\linewidth]{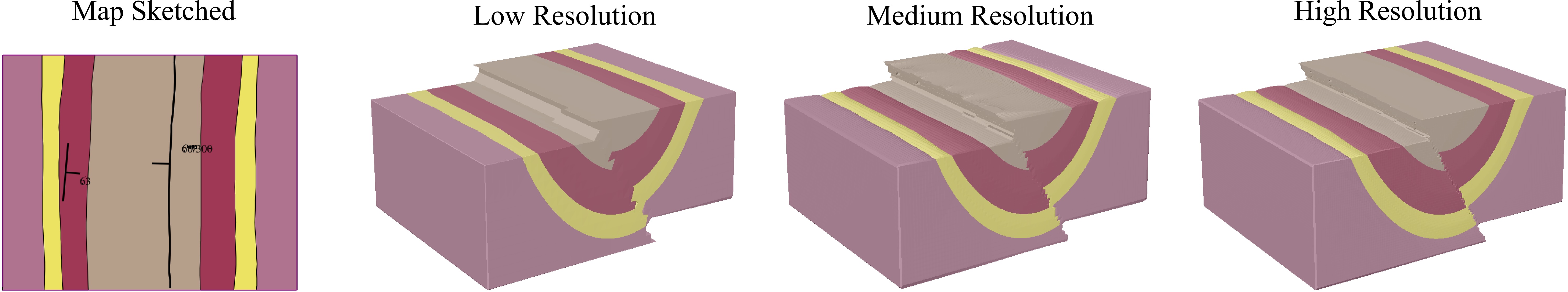}
  \caption{Different resolutions of the model created using Marching Tetrahedra.
  Low resolution models preserve the overall shape but are jagged in the fault region.}
  \label{fig:low_medium_high}  
\end{figure*}

Geo-Sketcher runs interactively on a Microsoft\textsuperscript{\textregistered} Surface Pro tablet PC (Intel Core i5 (3rd Gen) 3317U / 1.7 GHz with 4GB RAM) for creating models in low resolution.
A 3D model containing four different layers and one fault is generated in about $1.8$ seconds in low resolution, $14.1$ seconds in medium resolution, and $63.2$ in high resolution.
Where the resolution is the discretization of the space for sampling the implicits representing the geological model and creating the 3D model for visualization using marching tetrahedra.
Low resolution comprises a $55\times55\times20$ grid, medium resolution a $110\times110\times42$ grid, and high resolution a $180\times180\times70$ grid.
Fig.~\ref{fig:low_medium_high} presents the model described in all three different resolutions.
The difference of resolution in the presented example is more noticeable in the fault area, specially when comparing low to high resolutions.
All figures of the 3D models presented in the remainder of this section were captured using the high resolution to improve the visualization of the results.

The results shown in this section represent some simple examples of simple geological structures, to show that the Geo-Sketcher tool is able to reproduce some basic geology.
It also presents real world scenarios to show that it is capable of creating more complicated geology.

\begin{figure}[hb!]
  \centering
  \includegraphics[width=\linewidth]{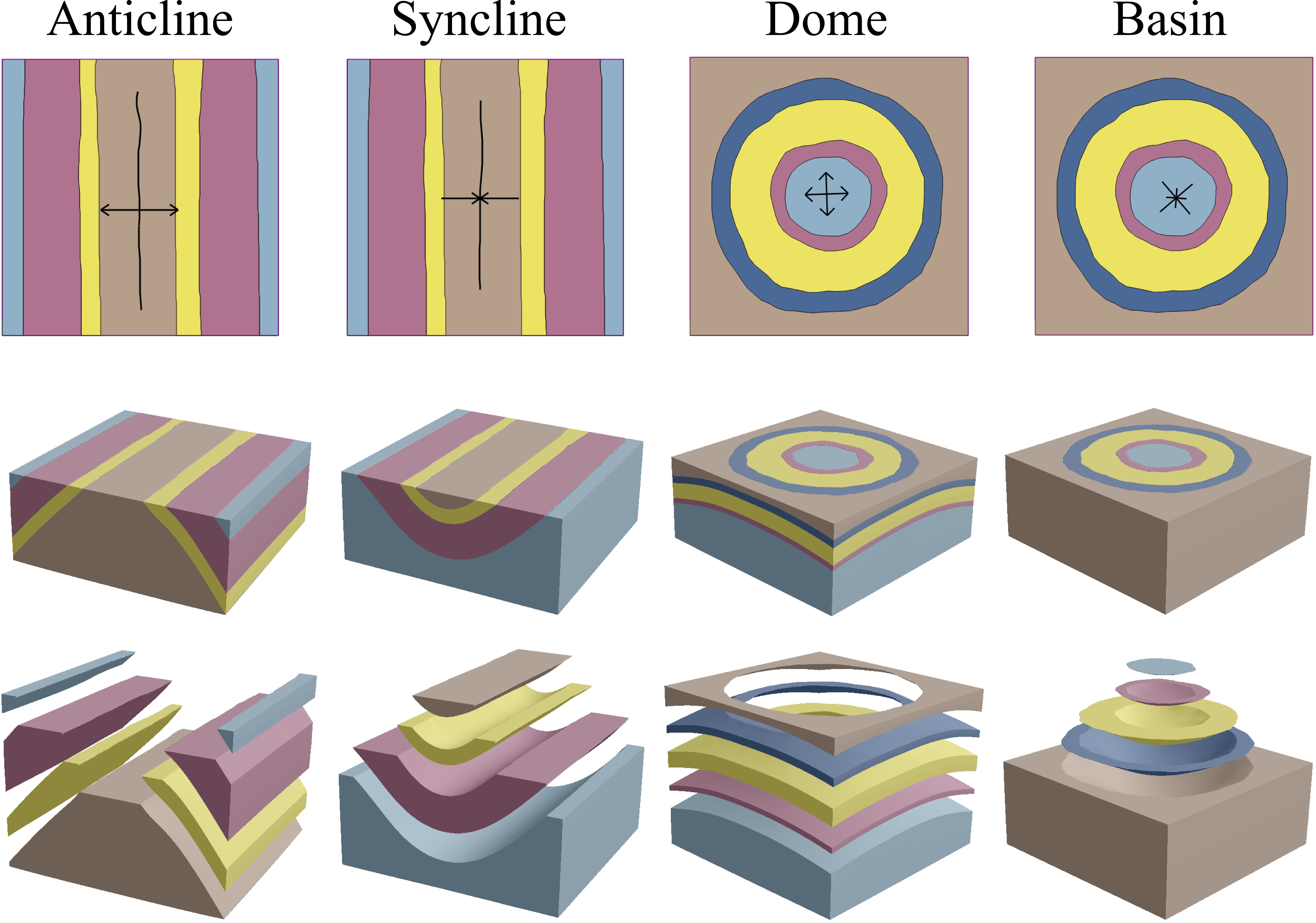}
  \caption{Results obtained by sketching basic fold structures using their specific symbols.
  First row presents the sketched map, the second row the generated 3D geological model, and the third row the same model in an exploded view.}
  \label{fig:basic-structures}  
\end{figure}

Our first experiment is to try to replicate the simple geological structures (folds and faults) defined in Section~\ref{sec:background}.
Fig.~\ref{fig:basic-structures} presents sketches of 4 different types of folds: anticlines, synclines, domes, and basins.
As it can be observed, the sketched map for anticlines and synclines, as well as for dome and basins, only differ by the geological symbols on them.
Based on this information the rock layers ages were automatically computed.
And all fold structures sketched were successfully converted into their respective 3D models.
Moreover, the automatic computation of the layers age makes it straightforward to create the exploded views of the models as shown in this section.

\begin{figure}[hb!]
    \centering
    \includegraphics[width=\linewidth]{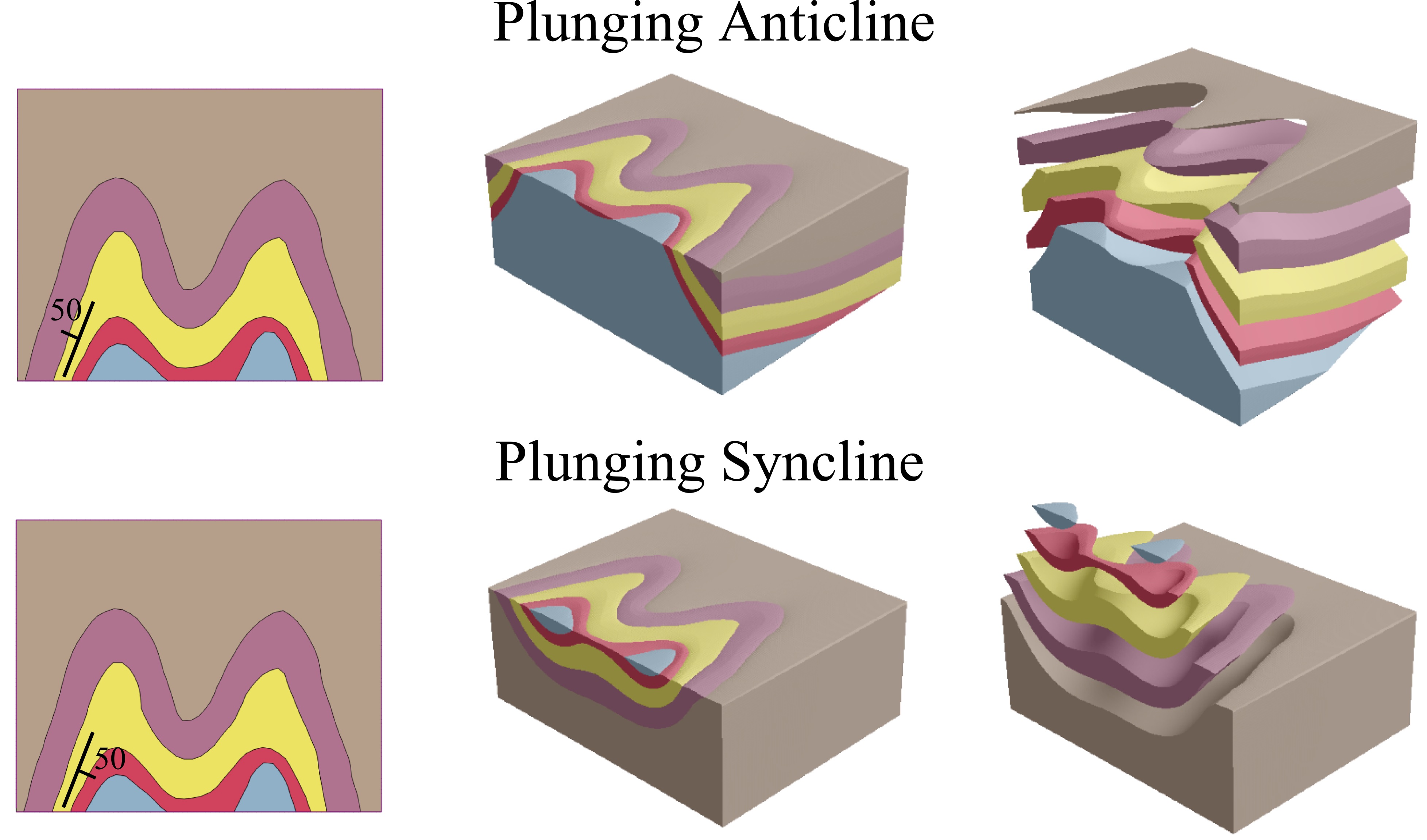}
    \caption{Sketching plunging anticline and syncline folds using a dip-and-strike symbol and \emph{W}-shaped contacts.}
    \label{fig:plunging}  
\end{figure}

\begin{figure}[ht!]
    \centering
    \includegraphics[width=\linewidth]{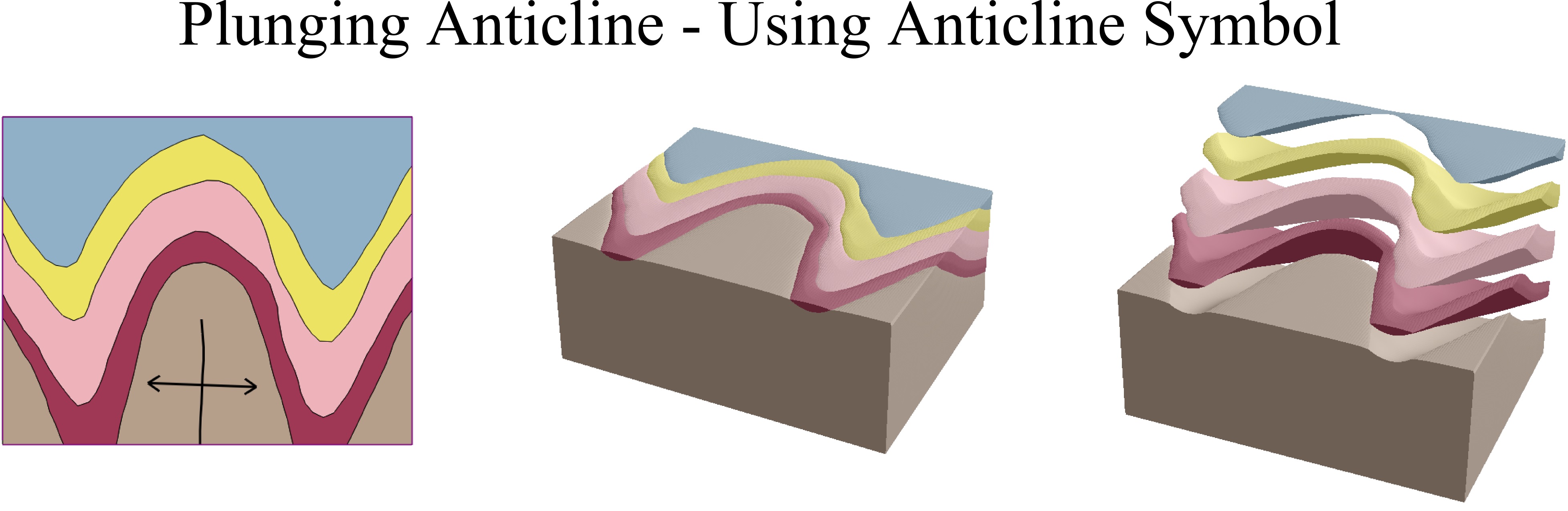}
    \caption{Sketching plunging anticline using anticline symbol and \emph{W}-shaped contacts.}
    \label{fig:plunging-symbol}  
\end{figure}

Plunging folds are interesting geological structures that can be observed in geological maps.
These types of folds are synclines or anticlines that have been tilted and after erosion they leave a \emph{V} pattern (or \emph{W} for two consecutive folds) when cutting the ground.
Fig.~\ref{fig:plunging} presents the construction of a model containing a plunging anticline and a plunging syncline. 
We do so by sketching the \emph{W} pattern and a single dipping symbol oriented according to the type of fold.
Fig~\ref{fig:plunging-symbol} also shows a plunging anticline but now created using the anticline symbol.

\begin{figure}[ht!]
  \centering
  \includegraphics[width=\linewidth]{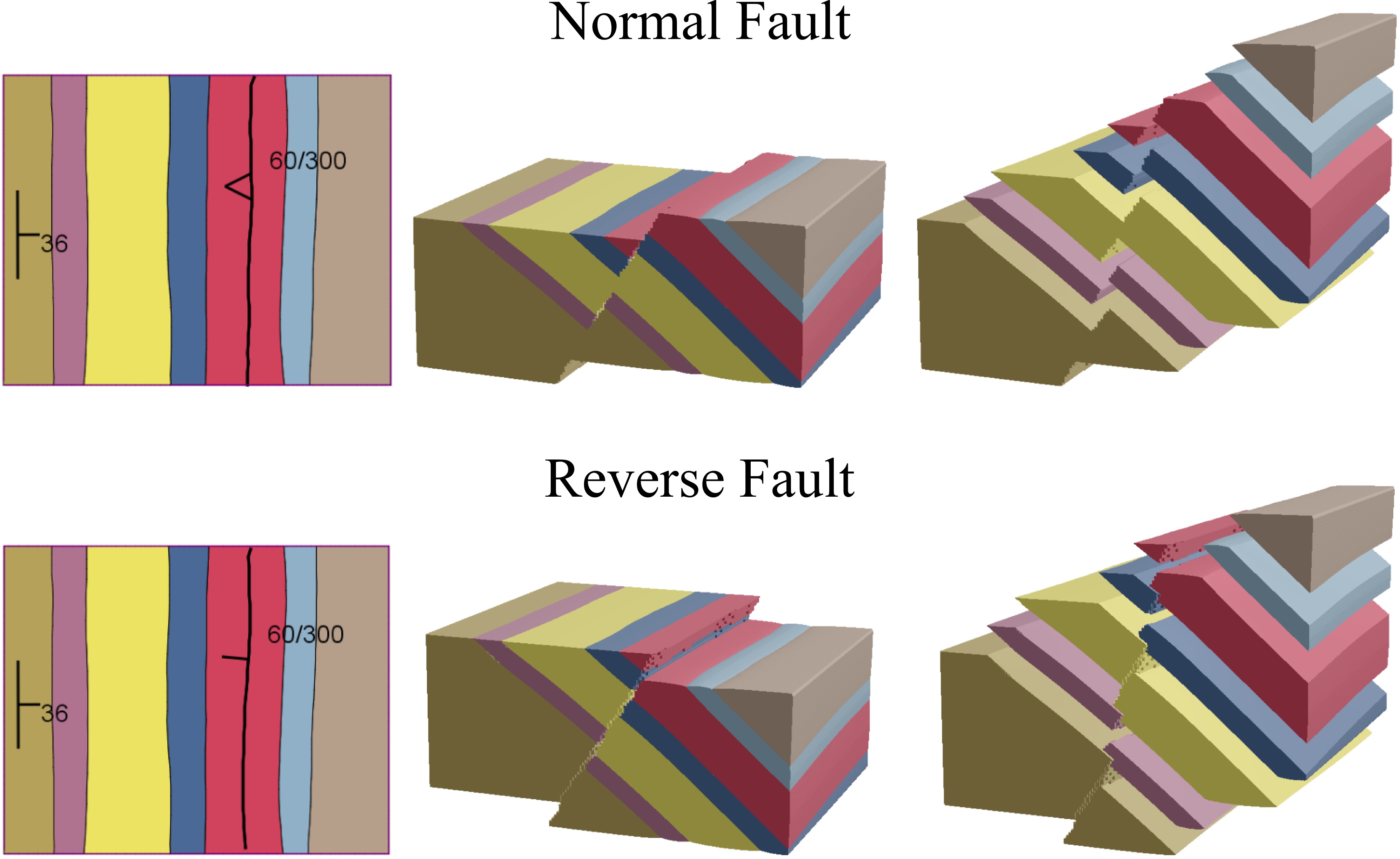}
  \caption{Sketch and 3D geological model constructed for normal and reverse faults.
  The dip-and-strike symbol is also sketched specifying the dipping of the rock layers.}
  \label{fig:faults-results}  
\end{figure}

Fig.~\ref{fig:faults-results} presents the sketch of normal and reverse faults, and the dip-and-strike symbol which were also described in Section~\ref{sec:background}.
In this example, we use the same layers with the same dipping for both normal and reverse faults.
As it can be observed, the layers dip approximates the amount defined by the dip-and-strike symbol.
Moreover, the normal-fault symbol generates a fault created by stress forces while the reverse fault symbol generates a fault created by compressional forces.
It is also possible to notice that the rock layers' thickness are well approximated since it maintains a constant thickness throughout the 3D model.

\begin{figure}[ht!]
  \centering
  \includegraphics[width=\linewidth]{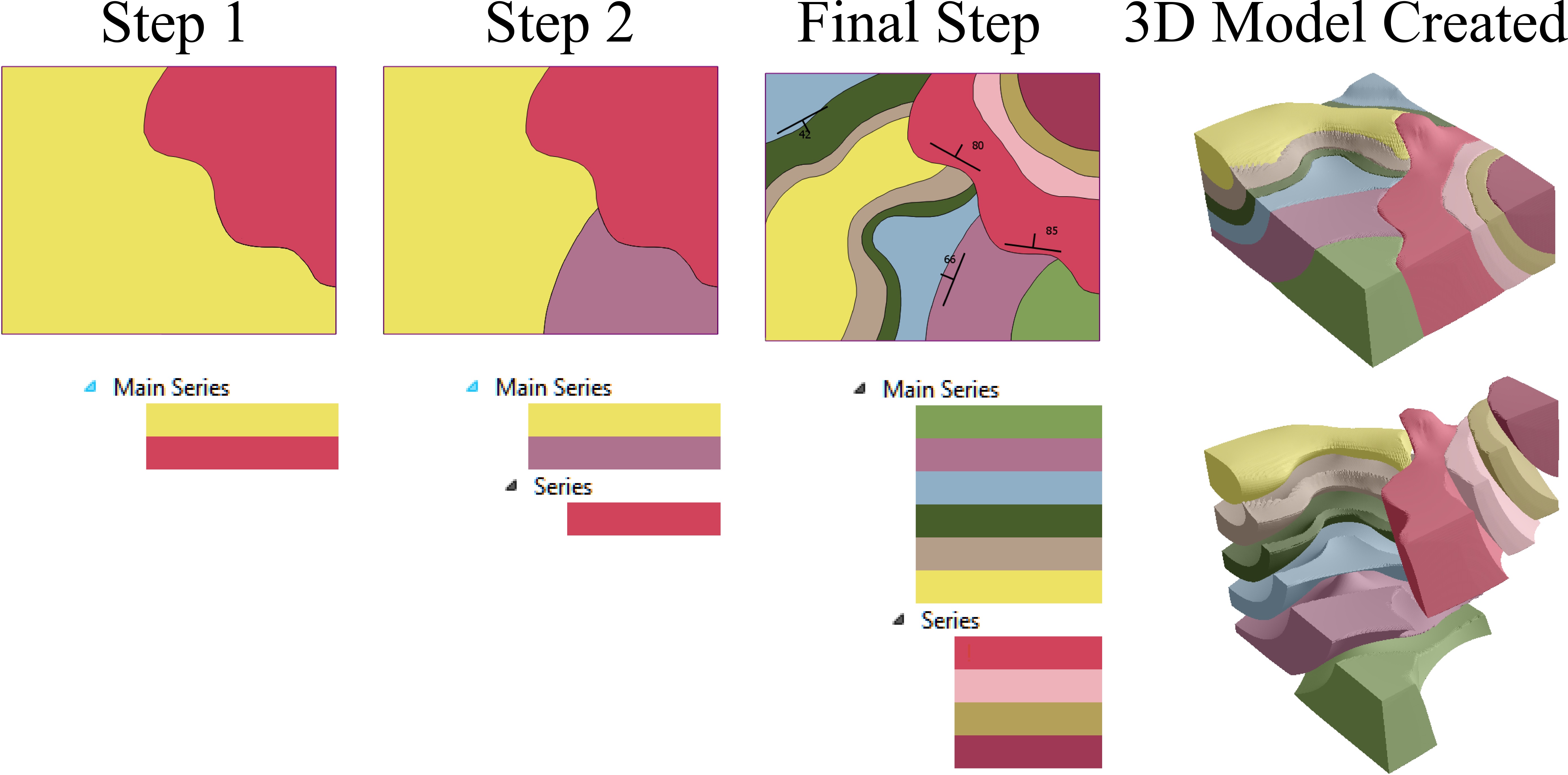}
  \caption{Creating geological model with an unconformity.
  In the first step the map sketched contains a single contact separating two rock layers.
  In the second step a new contact is sketched intersecting the previous one defining an unconformity.
  In the final map, more rock layers are added as well as strike and dipping symbols.}
  \label{fig:unconformity}  
\end{figure}

Fig.~\ref{fig:unconformity} shows an example of an unconformity being sketched and how the system automatically detects and properly create the new series and layers of rock.
The sequence of layers and series shown, go from older (top) to younger (bottom).
As it can be observed, the new series, defined by the unconformity contact, is subtracted of the older series, creating a valid 3D geological model with no overlap of rock layers.
According to our collaborators with expertise in geological modeling, creating even simple models like these ones can take hours, while using Geo-Sketcher the same task can be accomplished in less than a minute.

\begin{figure*}[ht!]
  \centering
  \includegraphics[width=0.8\linewidth]{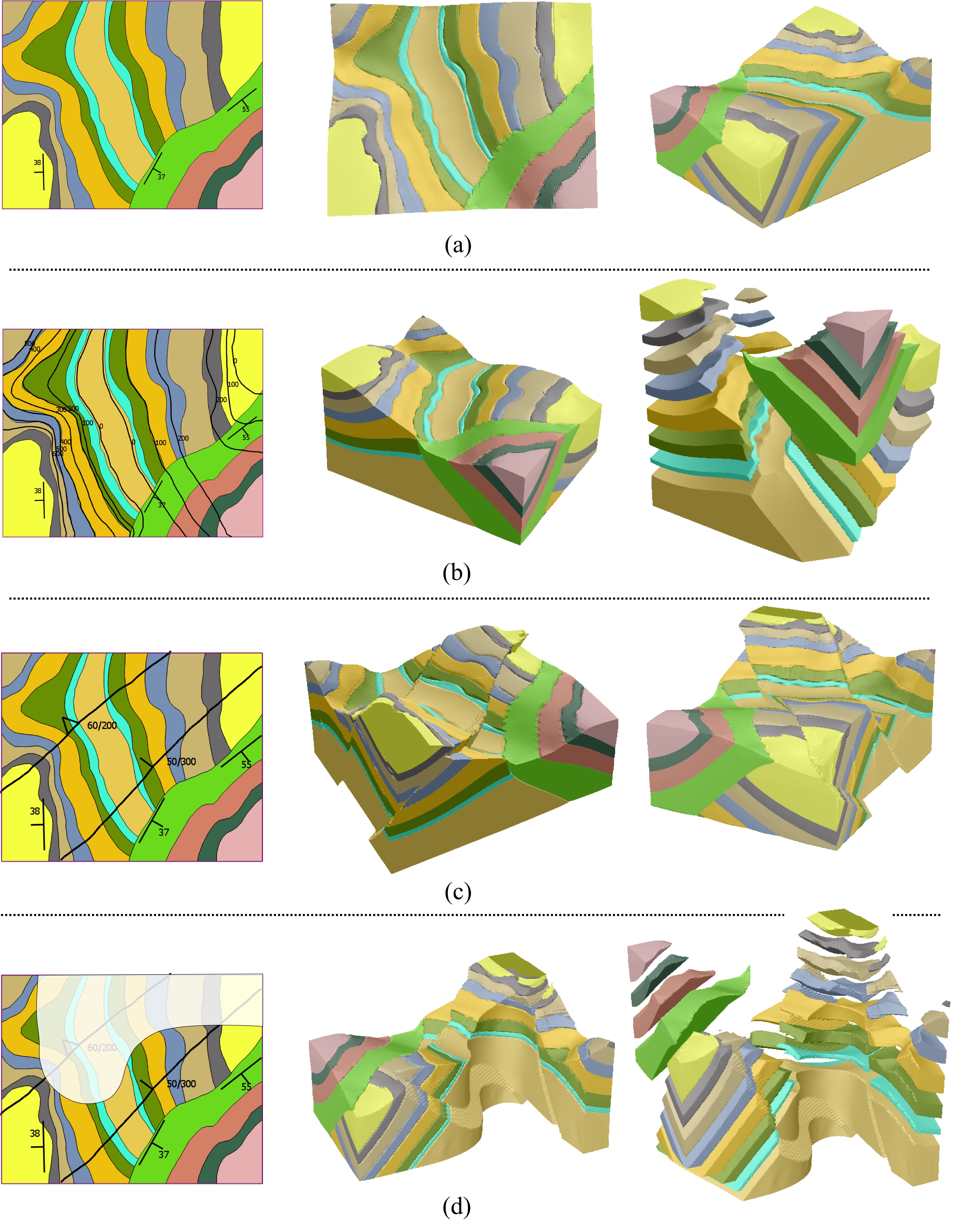}
  \caption{Complex model created using Geo-Sketcher.
  (a) sketched geological map and the 3D geological model constructed are shown.
  (b) same model as in (a) but now showing the topographic map and visualizing the 3D model on a different perspective.
  (c) a reverse and a normal fault are added to the model.
  (d) inspecting the interior of the model by sketching a cross-section curve.}
  \label{fig:synthetic}  
\end{figure*}

In Fig.~\ref{fig:synthetic} we present a geological model sketched with 12 different rock layers and an unconformity, on top of a sketched topographic map.
The model is shown before and after adding a normal and a reverse fault, and later cut using a cross-section sketch to show the interior of the model.

\begin{figure*}[ht!]
  \centering
  \includegraphics[width=0.8\linewidth]{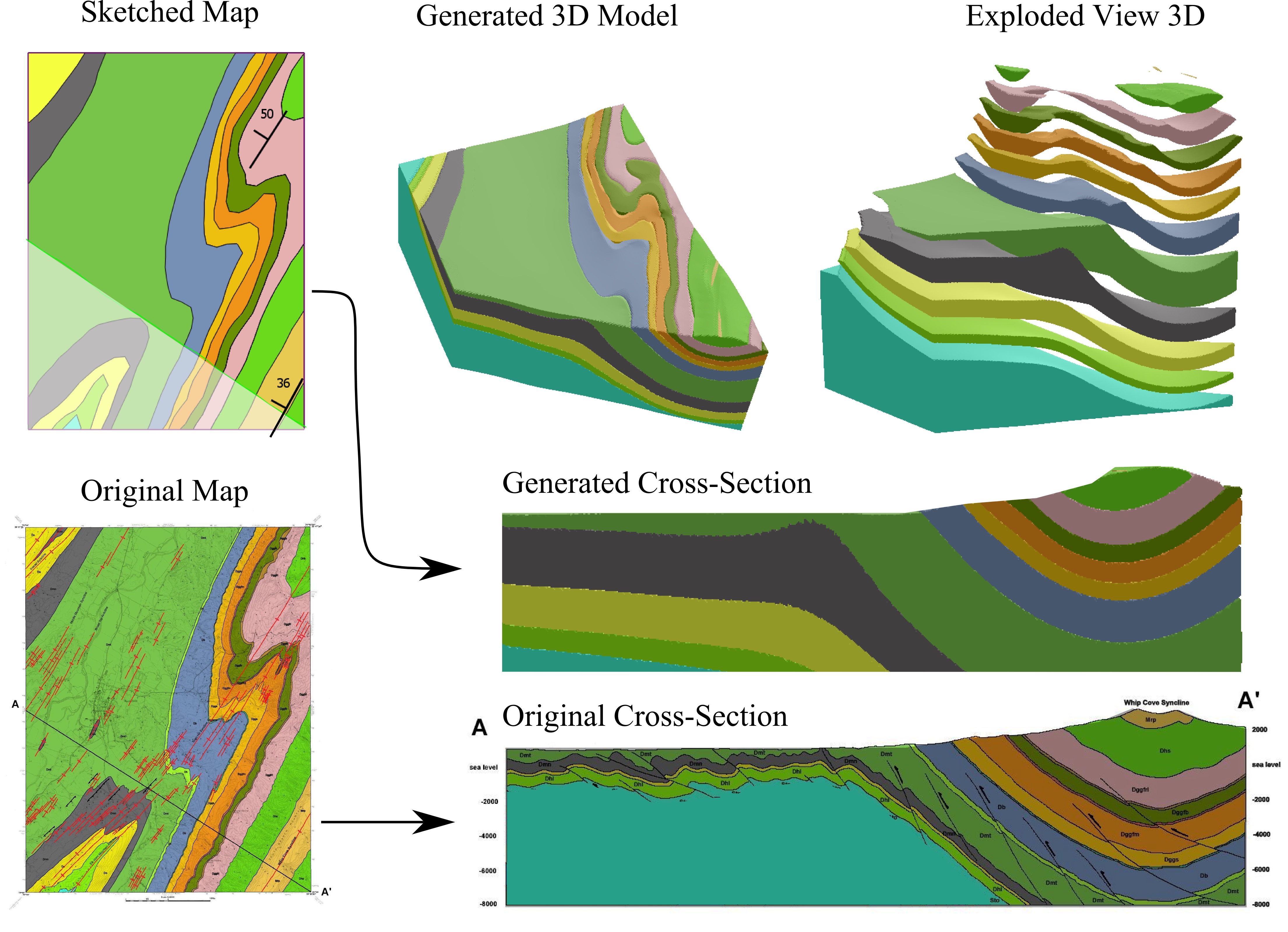}
  \caption{Sketched version of real geological map of the region of Moorefield, West Virginia, United States, compared to real map and cross-section~\cite{stuartLDean2001}.
  The overall shape of the constructed model is similar to the original one but there are some discrepancies in the layer thickness when compared to the existing cross-section.  
  }
  \label{fig:moorefield}  
\end{figure*}

The next example comes from a real world scenario (see Fig.~\ref{fig:moorefield}).
It is a simplified sketch of a geological map of the Moorefield quadrant in West Virginia, United States.
The original map comes with a cross section representation of the geological map as indicated by the line $A'A$.
The original cross-section was modified to hide some of the older geological layers not represented directly in the original geological map.
In this example we compare the results of the original map cross-section to the model generated by the sketched map.
It is possible to notice that the overall shape of the cross section is similar although it has some differences.
The first difference is the lack of some very thin layers in the original that were not sketched and therefore not presented in the constructed 3D model.
Another noticeable difference is the missing layer on the top of the hill in the original (yellow) which is very thin in the sketched model.
Finally, the thickness of some of the older layers (bottom) are also different from the original.
Most of the discrepancies are due to the simplification of the sketch, and the simple dip averaging when computing the layers' thickness.

\begin{figure*}[ht!]
  \centering
  \includegraphics[width=0.8\linewidth]{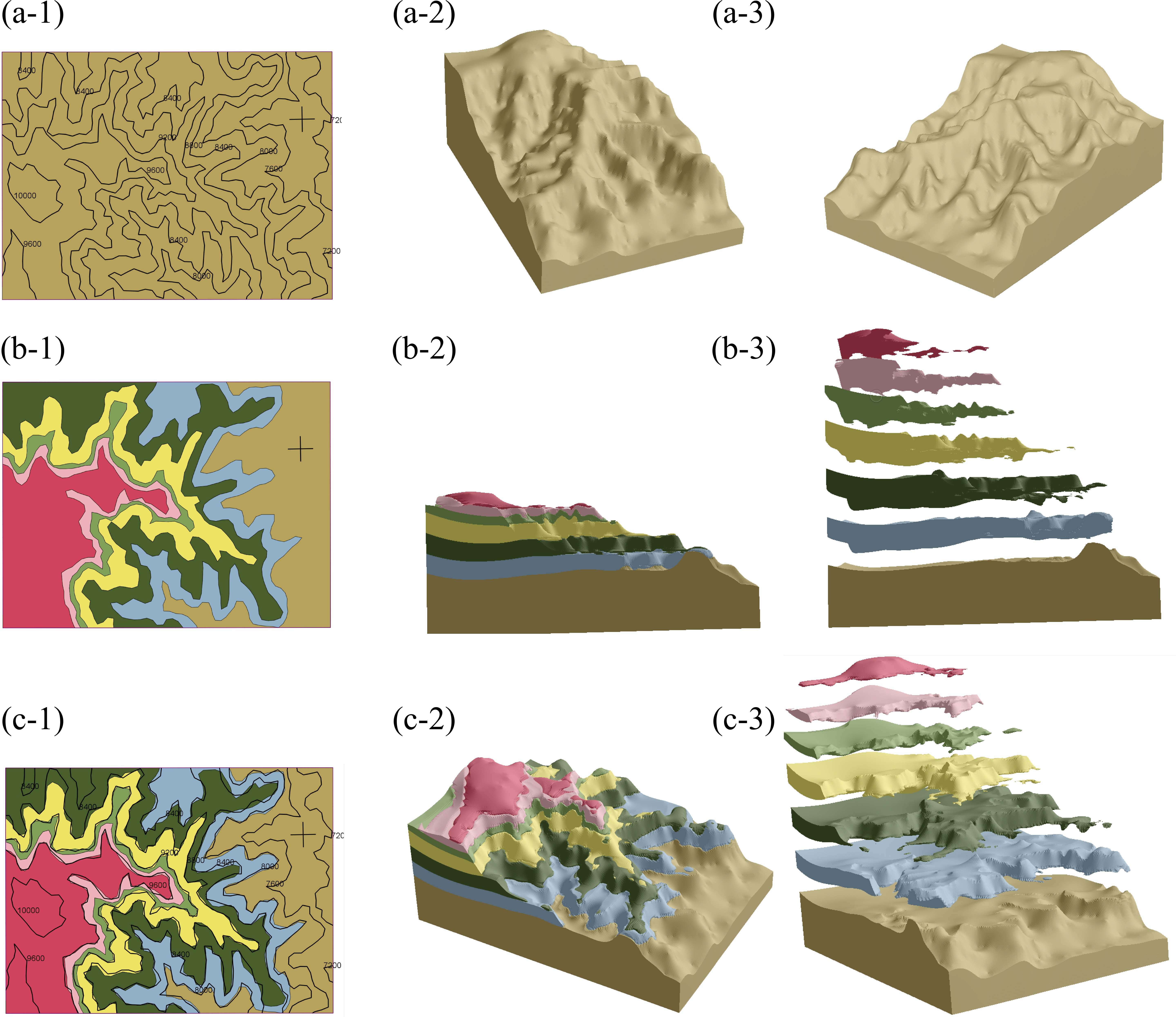}
  \caption{Geologic map of region of Utah, United States, sketched.
  (a) the sketch of the topographic map and the resulting relief created.
  (b) the geological map sketched using horizontal rock layers symbol and the resulting 3D geological model created.
  (c) showing topographic and geological maps with same results in a different angle.}
  \label{fig:utah}  
\end{figure*}

Finally, the last result is also a replication of a real geological map in Utah, United States (see Fig.~\ref{fig:utah}).
It is an example of horizontal layers and the use of the horizontal layers symbol combined with intricate topographic information.
First, the 3D terrain model is shown with the respective sketched topographic map.
Then the geological layers and the horizontal rock layers symbol is added and a complete 3D geological model of the region is created.
It is possible to observe that it approximates the horizontal layers although it has some minor discrepancies where the layers get too thin because of the topography.
These small differences are caused by the resolution of the 3D geological being constructed as well as some fluctuations of the height of the contacts by sketches not matching the exact height throughout the whole sketch.

We presented Geo-Sketcher to several industry experts, and the general feedback was that the tool has great potential to save hours of work.
Moreover, it was pointed out that our SBIM tool replicates common daily work activities where domain expert sketch and discuss different geological concepts and scenarios, given the interpretive nature of different stages of geological modeling.
Some experts suggested that Geo-Sketcher could also be used for teaching geological concepts to geology and reservoir engineering students, as well as a communication tool to explain geological scenarios to non-experts.
It was also reported that our tool can greatly improve the understanding of 3D block diagrams commonly used in geology (Fig.~\ref{fig:sketches}(b)) and improve 3D visual perception for domain experts
Finally, it was also indicated that the topographic sketch tool could add value to other areas of geology where contour curves are often used, such as mineral deposits maps. 



\section{Conclusions and Future Work}
\label{sec:conclusions-and-future-work}
The Geo-Sketcher represents several contributions for geological modeling and SBIM.
The choice of sketching of geological and topographic maps enables the expert to focus on the task of constructing the geological model rather than learning a new modeling tool or sketching language.
The automatic detection of the existing rock layers and their sequence, as well as the topographic contour curves automatic height computation and update, help expediting the model creation process.
Which gives the expert the ability to rapidly and easily create prototypes of geological models and try different geological scenarios.
The embedding geological rules in the sketching of geological and topographic maps and in the 3D model creation enables the construction of 3D geological models that are correct from the beginning avoiding problems if the model is exported for simulation, for instance.
Moreover, the ability to construct the 3D model \textit{on-the-fly} as the sketches are performed helps identifying mistakes early in the sketching process.
We also introduced in SBIM and geological modeling the use of HBRBFs, and presented with details how this mathematical tool can be used in such context.
And we have developed a different approach for creating faults in geological models that are more accurate than the ones previously described in the geological modeling literature.
Finally, the topographic map sketch itself can be used in other scenarios in geology, where contour curves are defined, as well as in other applications such as terrain generation for games.

Although our proposed system allows the rapid prototyping of 3D geological models with decent level of details, we believe that it can be enhanced by allowing other sketching perspectives other than top view.
Cross-sections are another form of representing geology.
They are side views of the geology and usually relies basically on lines representing faults and horizons.
Another sketching perspective could be directly in 3D, as proposed by Amorim et al.~\cite{amorim2012}, allowing adding some fine-tuned details to the model.
These three different views would be integrated such that changes in one perspective would be reflected in the other perspective.
The greatest challenge would be going from cross-section, or 3D view, back to geological-map view because of the geological map symbols.
When, where and how to translate the 3D shape into geological map symbols is an interesting problem on itself.
Sketch-based modeling of topographic maps can be extended to allow the modeling of terrains containing overhanging cliffs.
However, a new mathematical formulation would be necessary since the proposed RBF reconstruction does not support such features as described in Sec.~\ref{sec:rbf}.







\bibliographystyle{model3-num-names}
\bibliography{paper-bib}







\end{document}